\documentclass{jfm}

\usepackage{graphicx}
\usepackage{natbib}

\ifCUPmtlplainloaded \else
  \checkfont{eurm10}
  \iffontfound
    \IfFileExists{upmath.sty}
      {\typeout{^^JFound AMS Euler Roman fonts on the system,
                   using the 'upmath' package.^^J}%
       \usepackage{upmath}}
      {\typeout{^^JFound AMS Euler Roman fonts on the system, but you
                   dont seem to have the}%
       \typeout{'upmath' package installed. JFM.cls can take advantage
                 of these fonts,^^Jif you use 'upmath' package.^^J}%
      }
  \else
  \fi
\fi

\ifCUPmtlplainloaded \else
  \checkfont{msam10}
  \iffontfound
    \IfFileExists{amssymb.sty}
      {\typeout{^^JFound AMS Symbol fonts on the system, using the
                'amssymb' package.^^J}%
       \usepackage{amssymb}%
         \let\leq=\leqslant
         \let\geq=\geqslant
      }{}
  \fi
\fi

\ifCUPmtlplainloaded \else
  \IfFileExists{amsbsy.sty}
    {\typeout{^^JFound the 'amsbsy' package on the system, using it.^^J}%
     \usepackage{amsbsy}}
    {\providecommand\boldsymbol[1]{\mbox{\boldmath $##1$}}}
\fi

\newsavebox{\astrutbox}
\sbox{\astrutbox}{\rule[-5pt]{0pt}{20pt}}

\title[]{Transient Rayleigh-B\'enard-Marangoni Convection due to Evaporation : a Linear Non-normal  Stability Analysis.}

\author[F. Doumenc, T. Boeck, B. Guerrier and M. Rossi]
{F.\ns D\ls O\ls U\ls M\ls E\ls N\ls C$^1$ , \ns
 T.\ns B\ls O\ls E\ls C\ls K$^2$ , \ns
 B.\ns G\ls U\ls E\ls R\ls R\ls I\ls E\ls R$^1$\break
\and M.\ns R\ls O\ls S\ls S\ls I$^3$}

\affiliation{$^1$UPMC Univ Paris 06, Univ Paris-Sud, CNRS, UMR 7608, Lab FAST, Bat 502, Campus Univ,
Orsay F-91405, France\\[\affilskip]
$^2$Institut f\"ur Thermo- und Fluiddynamik, Technische Universit\"at Ilmenau,
Postfach 100565, 98684 Ilmenau, Germany\\[\affilskip]
$^3$UPMC Univ Paris 06, CNRS, UMR 7190, IJLRA, 4 place Jussieu, Paris F-75005, France}

\pubyear{2009}
\volume{??}
\pagerange{??--??}
\date{?? and in revised form ??}

\begin{document}

\maketitle


\begin{abstract}

The convective instability in a plane liquid
layer with time-dependent temperature profile  
is investigated by means of a general  method suitable 
for linear stability analysis of an unsteady basic flow. The method is based
on a non-normal approach, and  predicts the onset of instability, critical 
wavenumber and time. 
The method is  applied to transient Rayleigh-B\'enard-Marangoni  convection 
due to cooling  by evaporation.  
Numerical results as well as theoretical scalings for the critical parameters as function of 
the Biot number are presented  for 
the limiting cases of purely buoyancy-driven and purely surface-tension-driven  convection.
Critical parameters  from calculations are in good agreement with those from
experiments on  drying polymer solutions,
where the surface cooling is induced by solvent
evaporation.

\end{abstract}



\section{Introduction}
\label{sec:intro}

\noindent  Thermally driven flows near liquid interfaces continue to be
an active area of research since the first experimental studies by H. B\'enard
over 100 years ago. They are characterized by a multitude of interacting
physical mechanisms and display a large variety of regular and complex flow
patterns. Experimental and theoretical studies of such flows have
stimulated and  accompanied  the development of the theory of pattern
formation and nonlinear phenomena in general. Recent developments are
summarized in a number of monographs and review papers
(see for example \cite{Colinet:2001,Nepomnyashchy:2006,Bodenschatz:2000}).

\noindent The driving forces in these thermally driven flows
are surface-tension gradients (Marangoni forces) and buoyancy  forces,
which originate from the temperature dependency of surface tension and density, respectively.
Both mechanisms are classically studied
in convection taking place  within a  finite layer  subjected to a  steady   temperature difference:
 for the buoyancy-driven case, this is  the celebrated Rayleigh-B\'enard convection
 formulated by \cite{Rayleigh:1916}, for the surface-tension driven case
this is the B\'enard-Marangoni convection studied  by \cite{Pearson:1958}.  In both cases, a certain critical temperature difference must be
applied in order to sustain convective motion. The  theoretical predictions for such critical
temperature differences are in excellent agreement with experiments (\cite{Chandrasekhar:1961,
Schatz:1995}). Thermal convection can also appear  when the external conditions
are time-dependent, e.g. by  a modulation or abrupt change of
cooling or heating. In this case, the
basic conductive  temperature distribution
is time dependent, and the stability problem becomes a non-autonomous one,
i.e. the solution cannot in general be sought in the form of
exponentials $\exp(\sigma t)$. For time-periodic modulation of the basic
state one can resort to Floquet theory (\cite{rose71,Bhadauria2002}) but, in the general case, the
formulation and prediction of
critical conditions for the onset of convection becomes
much less clear-cut than for a steady base state.
The following two basic approaches are common for a general,
time-dependent basic state:
\begin{itemize}
\item reduction to an autonomous problem by the  {\em frozen-time
assumption}, whereby the basic state is supposed
to evolve much more slowly than the perturbations, and
\item solution of the full non-autonomous linear perturbation problem  for some
 initial conditions which are supposed to be representative. This is called {\em amplification theory} (denoted by AT in  the following).
\end{itemize}
Both approaches go at least back to the 1960s. The first one was used
by \cite{Lick:1965} and  \cite{Currie:1967}  and the second by \cite{Foster:1965}.
Later work by \cite{Homsy:1973} introduced energy stability ideas, but the resulting bounds are not necessarily
useful for predicting instability thresholds and the mathematics is
considerably more involved. In each of these works the focus was on
buoyancy-driven convection.

\noindent  The frozen-time approach was  used  for the Marangoni convection by
\cite{Kang:1997}.  These authors studied the dynamics  of  a fluid layer 
subjected to  a sudden change in surface temperature.  However the frozen-time  method  was  
only  applied  for late times in the evolution.  For early times
they formulated the so-called propagation theory, whereby the problem is again
reduced to an autonomous one by using a certain
similarity solution for  base state and perturbations. This approximation is actually 
appropriate for an infinite layer.   The   frozen-time model   relies on the validity of  the assumption of fast instability
development (relative to the base state). This could be inadequate  at instability
thresholds, at which  the instability can develop on the same time scales as the
base state. The other method namely  the amplification model,  has a sounder mathematical
foundation, but comes with the necessity of identifying representative initial
conditions  and amplification levels for  the perturbations.

\noindent  A specific analysis suitable for  transient problems is
developed in this paper, which is an extension  of the  previous
amplification model.  The underlying concept    is  called the
non-normal approach.  In such a  case, the choice of  initial
perturbations is not somewhat  arbitrary as in AT but it is guided
by an optimization technique.  This procedure  consists in
identifying, for a given time $t$ and a given wavenumber $k$, the initial
perturbation with the strongest amplification (called  the optimal
perturbation). Hence this method may exhibit initial conditions
which  are not a priori obvious configurations. As a consequence
it gives the maximal value for the  perturbation norm that was
previously  defined for the problem under study. This cannot be
achieved through an AT simulation. Concerning the determination of
initial conditions,
we note that the non-normal approach has also
been  used   for instability problems of steady base flow, e.g.
in parallel shear flows, such as boundary layers
(\cite{Schmid2001}) or plane Poiseuille flow~(\cite{Reddy:1998}).
For such problems, there was a long-standing  gap between the
standard two-dimensional
 Tollmien-Schlichting modes  obtained using the
classical normal mode analysis -- though these modes were only
observed in some very controlled experiments -- and the general
experimental observation of streak generation. It is by resorting
to non-normal mode analysis that  the coherent structures observed in
boundary layer transition, i.e. streamwise rolls and streaks,
could be deduced from an optimal approach and their generation
linked to  the lift-up mechanism proposed by Landahl
(\cite{Schmid2001}).

\noindent The non-normal approach can be extended  to transient
cases. This is  precisely  what we propose here. Such a  method   provides the
strongest amplification, i.e. the optimal growth at  a given
duration $T$ after initial conditions have started their
evolution. If this amplification is sufficiently large,  the
optimal  perturbations may result in the modification of the basic
flow at time $T$ and we can speak of an unstable regime in a way to
be defined by a norm.  To the best of our knowledge, this approach
is the only  one capable to provide clear-cut answers on
instability problems for truly unsteady basic flows.  However, as
mentioned by one referee, this definition of the stability
condition is   blurred  owing to a  certain arbitrariness in
choosing the norm and defining  the critical amplification gain.
This is inherent  to any unsteady  linear stability problem and does not
depend on the particular method. It is then more suitable to view
the results as the estimation of a transition region between a
domain exhibiting  strong convection and a domain where initial
perturbations are damped or have no time to significantly develop
during the transient regime. The non-normal approach allows one to
characterize this transition domain properly. The results
presented here for different norms and different critical
amplification values show that this transition region is thin, so
that the notion of stability threshold is still valid for this
transient problem.  On the contrary, the frozen-time assumption
may fail, for instance  if the base state evolves on the same
time scale  or faster then the unstable modes characterizing its
frozen-time stability.  In that case, the computed growth rates might
not correspond to any true amplifications, at the system time scale.
This might affect the determination of critical 
conditions~\footnote{ In the frozen-time assumption,  the critical 
conditions are determined  using  the quasi-static growth rates $\sigma(t)$. 
However the pertinent character of this method   stongly  depends  
on the rapid variation of these ``quasi-static growth rates''  
with the basic state changes. For instance, the amplification between times $t_1$ and $t_2$ would be,  at zeroth order WKB theory,  equal to 
$ \int^{t_2}_{t_1} \sigma(t) dt$. So if the instantaneous growth rate strongly varies during this interval, 
its positive value at a given moment may  not be interpreted in the right way.   }.

In the present work,
we  provide the results obtained by means of the non-normal
approach as well as the frozen-time approach. In the specific flow
case presented here, the quasi-static method  is shown to  provide
similar results except for some  particular quantities.

\noindent The proposed analysis is general and can be easily extended to many other unsteady problems, e.g.,
chemically driven hydrodynamic instabilities (\cite{Eckert2004}).   In the present paper, we  determine  the onset of convection
in  a drying experiment, which leads to  an unsteady B\'enard-Marangoni problem.
More specifically we study  the sudden cooling  due to evaporation of a liquid layer,
where the decrease of surface temperature is induced by the vaporization latent heat.
This problem has been the subject of many experimental or theoretical studies
(see for example \cite{Berg66}, \cite{Vidal68},
and more recently \cite{Mancini04} or \cite{mous04}).
The specific motivation for our work  is the drying process  studied in
experiments by~\cite{tous08} performed on a polymer solution (Polyisobutylene/Toluene).
The solution initially at the ambient temperature is poured
in a dish located in an extractive hood. When evaporation of toluene begins,
convective patterns are observed at the very beginning of
the experiment (quasi-instantaneous or less than 100~s after
pouring the solution). They disappear well before the end of the
drying.
The very large Lewis number $Le=\kappa/D_{mol}$  ($\kappa$  and $D_{mol}$ denote  the thermal and mass diffusivity) of the polymer solution (about $10^{3}$) 
is an indication that the thermal diffusion is faster and  that convective patterns observed in the first
minutes should be mainly driven by thermal effects. Two experimental observations detailed in \cite{tous08} support this thesis.
First, a few experiments were conducted with deuterated solvent, whose density is higher than the polymer density.
In that case, the density of the solution decreases when the polymer concentration increases,
leading to a stable situation if the solutal Rayleigh-B\'enard problem is considered.
Since no differences were found with the experiments conducted with the standard solvent,
we can exclude solutal buoyancy as a dominant mechanism.
Second, free surface temperature fields measured by infrared camera showed that the end of free convection
was related to the duration of the transient thermal regime.
In these free convection experiments it can be inferred  from the
work on steady convection by \cite{Pearson:1958}, that the
Marangoni effect is dominant for thicknesses typically less than
1cm and the buoyancy dominant for higher thicknesses. For a more
accurate description of  the experiments   see~\cite{tous08}.

\noindent  The   paper is organized as follows.  In section \ref{sec:equations},
we present the basic  assumptions of the model and  the governing equations.
Thereafter the unsteady basic state is described  and a specific  stability analysis is introduced in section \ref{linearoptimal}.
In  particular the non-normal method is explained  and the choice of   norms  is discussed. Section \ref{results} contains the main numerical findings for the two limiting situations:  the pure Rayleigh-B\'enard   and the pure Marangoni case. Critical conditions  for the optimal
modes are presented. Part of  these numerical results are also obtained by a  scaling analysis. A  comparison of this method with the frozen time approach is also performed. Finally, in section  \ref{Experiment},  the comparison  with  experimental results is discussed.

\section{Mathematical Model}
\label{sec:equations}

\subsection{Basic Assumptions of the Model}

\noindent The mathematical model of Rayleigh-B\'enard-Marangoni
convection used throughout this paper  is based on a one-layer
model  in which    three assumptions are made :   (1) the    upper
surface remains planar, (2)  the layer thickness $d$ remains constant,
(3) the heat and mass fluxes  across  the upper surface are given
by transfer coefficients.   Moreover   our analysis is restricted
to fluids  characterized by a Prandtl number $Pr =\nu/\kappa  \geq 1 $ with   $\nu$  the  kinematic viscosity. This
turns out to be  the case of most  liquids (including water and
organic solvents). In that instance, the  thermal diffusion time
scale is always larger than  the viscous diffusion time scale.

\noindent  The first hypothesis can be tested as follows. Two
modes of instability are known to occur in the B\'enard-Marangoni
problem  (\cite{scri64,reic81,gous90}). One is generated from the
interaction of the velocity perturbations with the basic
temperature, while the other mode, characterized by a  wavelength
long compared with the layer  thickness, is due to the coupling of
the Marangoni effect with the deflection of the free surface.
Mathematically,  surface deformation can be neglected  on scales
of order $d$ when the Laplace pressure associated with a curvature
$1/d$ is large compared with the dynamic pressure.  This condition
(\cite{Davis87})   corresponds with the smallness of the
crispation number $Cr \equiv  {(\rho \nu \kappa)}/{(\sigma d)}$ where $\rho$ and $\sigma $
 respectively denote   fluid density and  surface tension.
Moreover, the Galileo number $Ga \equiv  {(g d^3)}/{(\nu\kappa)} $
characterizes the relative importance  of gravity ($g$  gravity constant) and   diffusion.
A large value of the Galileo number indicates that gravity
stabilizes the long-wave mode.
The free surface deformation can be  neglected if $Cr \ll 1$ and $Ga \gg 1$. Such conditions  are shown  be
satisfied for the experiments considered here.

\noindent The second hypothesis can be adopted  if   $Pe \ll 1$,
where  the Peclet number $Pe \equiv {(d{v}_{ev})}/{ \kappa} $ 
is defined as the ratio between 
${v}_{ev}$ the interface
velocity  due to evaporation which is equal to minus the time
derivative of    $d(t)$, and the thermal velocity scale $ \kappa / d$. 
Indeed  when $Pe \ll 1$,  the surface
displacement  $ {v}_{ev} \delta_{diff}  $ remains negligible
compared to the total thickness $d $ during  the problem
characteristic time i.e. the diffusion time  $ \delta_{diff}
\equiv  {d^2}/{ \kappa}$.  In the  experiment~(see  section
\ref{Experiment}), the Peclet number is smaller than  $0.1 $.

\noindent Finally let us discuss   the third assumption.  The boundary condition at the free surface results from the coupling between
the system and its surroundings. In evaporation experiments, the
solvent flux and thus the temperature gradient in the fluid
depends on the heat and mass transfer with the ambient air.
Several authors have developed numerical or theoretical studies
taking into account this coupling, see for example
\cite{merk03,Colinet:2003,ozen04,mous04}. In this paper we adopt a simple description   by
global heat and mass transfer coefficients, as our main interest is directed on the
transient character of the problem under study and not on the detailed description of the transfer per se.

\subsection{Governing equations}

\noindent  We  formulate the basic equations  in
a Cartesian coordinate system, where the   bottom of the layer
coincides with the plane $z=0$ and the upper free surface
with $z=d$. 
The fluid is  characterized by a
density $\rho$, a kinematic viscosity $\nu$, a thermal diffusivity $\kappa$  and a
thermal expansion coefficient $\alpha$.  The Boussinesq approximation is assumed to govern  the velocity field   $\boldsymbol{v}=v_x \boldsymbol{e}_x+v_y \boldsymbol{e}_y+v_z \boldsymbol{e}_z$   and  temperature field $T$
\begin{eqnarray}
\label{model-nsdimensional}
\frac{\partial \boldsymbol{v}}{\partial t} + (\boldsymbol{v}\cdot \nabla)\boldsymbol{v} &=& -\frac{\nabla p}{\rho}
+ \nu \nabla^2 \boldsymbol{v}  + g \alpha (T - T_{\infty}) \boldsymbol{e}_z,~~~~  \nabla \cdot\boldsymbol{v} = 0,\\
\label{model-heatdimensional}
\frac{\partial T}{\partial t} +
(\boldsymbol{v}\cdot \nabla)T &=&
 \kappa \nabla^2 T,
\end{eqnarray}
 where $g$  denotes the acceleration due to gravity and  $ T_{\infty} $ the temperature of the ambient air.
In this  approximation, the density $\rho$ is taken to be the density at $ T=T_{\infty} $.   At the    bottom,
the velocity satisfies the no-slip condition and the wall is assumed
 adiabatic since, in the experiment,  the bottom of the dish is thermally insulated by an air gap.
\begin{equation}
\label{model-bcbottomdimensional-noslip}
\boldsymbol{v}=0, \hskip 1cm  \partial_z T=0 \hskip5mm\mbox{at $z=0$.}
\end{equation}

\noindent  The  upper boundary conditions  are more involved. Assuming a planar surface, 
the local evaporation mass flux reads
\begin{equation}
\label{chirie1}
 Q_m(T_s(x,d,t))=\rho(v_z(x,d,t)-\frac{d}{dt}(d))    
 \end{equation}
Moreover  the global mass balance reads  (no fluid is introduced to counterbalance the evaporation mass loss)
\begin{equation}
\label{chirie2}
 \bar{Q}_m=-\rho\frac{d}{dt}(d) =\rho v_{ev}
\end{equation}
where $\bar{Q}_m$ is the mean evaporation flux over the free surface. From relations \ref{chirie1}
and \ref{chirie2} we get :
\begin{equation}
\label{chirie3}
 v_z= \frac{Q_m(T_s)- \bar{Q}_m}{ \bar{Q}_m} v_{ev}
\end{equation}
If we assume that the flux variations $Q_m(T_s(x,d,t))- \bar{Q}_m$  are much smaller than the flux $ \bar{Q}_m$ itself, 
then $ |v_z| \ll  v_{ev} \equiv Pe ~ \kappa / d $,   the Peclet number being   defined  using the evaporation velocity (see section 2.1).
Since $Pe \ll 1$ and $\kappa / d $ is the velocity scale, 
the kinematic boundary condition so reduces  to
\begin{equation}
\label{model-vztopzero1}
v_z=0 \hskip5mm\mbox{at $z=d$.}
\end{equation}
In addition, the balance of tangential forces at the upper surface
requires that the velocity field satisfies
\begin{equation}
\label{model-marangoni1}
\rho \nu \partial_z v_j= \partial_j
\sigma(T_s), \hskip5mm (j=x,y)   \hskip5mm\mbox{at $z=d$,}
\end{equation}
where $T_s$ denotes the   temperature inside the fluid layer at
$z=d$ and  $\sigma(T) $ denotes  the surface tension  which is
assumed to be   a linearly decreasing function of temperature,
\begin{equation}
\label{model-sigmaoftemp}
\sigma (T) = \sigma (T_{\infty}) - \gamma (T - T_{\infty}),~~ ~~  \gamma \equiv -\frac{d\sigma}{d T} >0.
\end{equation}

\noindent  Finally, the conservation of energy flux should be imposed at the interface. It reads :
\begin{equation}
\label{model-alphamodel0}
 -\lambda \frac{\partial T}{\partial  z} +  h_{th}(T_{\infty} -T_{s})  ={L}Q_{m}(T_{s})
\end{equation}

\noindent  The first l.h.s. term represents the heat conduction in
the liquid, $\lambda$  denoting the thermal conductivity related
to the thermal diffusivity {\it via} $\lambda = \kappa \rho C$
with $C$ the liquid specific heat. The second l.h.s. term
expresses the heat flux density in the gas using a simple
phenomenological model based on the heat transfer coefficient
$h_{th}$ and the difference between the  air temperature $T_{\infty}$ far from
the liquid  and the surface temperature $T_s$. Finally
the cooling effect due to solvent vaporization is expressed in the
r.h.s. term, where $L$  stands for the latent heat of vaporization
and $Q_{m}$ for the   solvent mass flux per unit area. This latter
quantity depends on the surface temperature $T_{s}$.  In the
framework of the one-layer model,  one imposes
\begin{equation}
\label{Qm}
Q_{m}(T_{s})=  h_{m}(c_{s}(T_{s})-c_{\infty} )
\end{equation}
where $c_{s}$ (resp. $c_{\infty}$) denotes the  solvent  concentration  in  the gas phase  near the surface (resp. far from the surface) and $h_{m}$ is the phenomenological
mass transfer coefficient in the gas. Assuming local thermodynamic equilibrium, $c_{s}$ directly depends on the surface temperature through the saturated vapour pressure.
The variable $Q_{m}$   can be  linearized  around $T_{\infty}$ which leads to the final expression:
\begin{equation}
\label{model-alphamodel}
-\lambda \frac{\partial T}{\partial  z} +  H_{th}(T_{\infty} - T_{s}) = {L}Q_{m}(T_{\infty}),~~ ~~  H_{th}= h_{th} +{L} \frac{\partial Q_m}{\partial T} (T_{\infty}).
\end{equation}

\noindent Initially the liquid layer is  isothermal  i.e.  $T(z,t=0)= T_{\infty}$.   At large times,  the system reaches a steady state   in which the temperature in the layer is again  uniform  but  with a temperature difference $  T_{\infty}-T_{s}= \Delta T_{st}   >0 $ with the gas located far from the surface. This difference is   imposed by  the condition (\ref{model-alphamodel}),     where
\begin{equation}
\label{DeltaTsdefinition}
\Delta T_{st} \equiv   \frac{L Q_{m}}{ H_{th}}.
\end{equation}
 \noindent To put the above equations  in a non-dimensional form, scales for temperature, length, velocity and time are needed.
The temperature  difference $\Delta T_{st} $  provides the  temperature  scale  and the   layer  thickness $d$
 the relevant length scale.   Two  velocity scales can be introduced   in this problem namely the viscous velocity scale $V=\nu/d$ and the
thermal velocity  scale $V=\kappa/d$.  Here   only   the  thermal scale $V=\kappa/d$ is used. Finally the appropriate time scale is the thermal diffusion
time $d/V={d^2}/{ \kappa} $.
The non-dimensionalization  leads to equations  for $\boldsymbol{v}$ and  $\theta(z,t) = ({T(z,t)-T_{\infty}})/{\Delta T_{st}} $ :
\begin{eqnarray}
\label{model-nsthermal}
\hskip-1.5cm& & \frac{1}{Pr} ( \partial_t\boldsymbol{v} +
  (\boldsymbol{v}\cdot \nabla)\boldsymbol{v}) = -\nabla p
+    \nabla^2 \boldsymbol{v} +  Ra \theta \boldsymbol{e}_z
,\\
\hskip-1.5cm& & \nabla \cdot\boldsymbol{v}  =  0,\\
\label{model-heatthermal}
\hskip-1.5cm & &  \partial_t \theta + (\boldsymbol{v}\cdot \nabla)\theta  =  \nabla^2 \theta  , \\
\label{model-bctopthermal}
\hskip-1.5cm& &  \label{bcondBiothermal1}\partial_z v_x + Ma\, \partial_x
\theta=\partial_z v_y + Ma\, \partial_y \theta=0  \hskip7mm\mbox{at $z=1$}.\\
\hskip-1.5cm& &  \label{bcondBiothermal2} \partial_z \theta +Bi\, \theta +Bi=0  \hskip4mm\mbox{at $z=1$}, \hskip7mm \partial_z \theta=0 \hskip2mm\mbox{at $z=0$}, \\
\hskip-1.5cm& &  \label{bcondBiothermal3} v_z=0, \hskip2mm\mbox{at $z=1$} \hskip5mm  v_x=v_y=v_z=0 \hskip5mm\mbox{at $z=0$}.
\end{eqnarray}
in which  the    Rayleigh, Marangoni, Prandtl and Biot  
 numbers 
\begin{eqnarray}
  Ra =  \frac{\alpha g d^3 \Delta T_{st}}{\nu \kappa}, \hskip2mm
 Ma  = \frac{\gamma d \Delta T_{st}}{\rho \nu \kappa},\hskip2mm  Pr=  \frac{\nu}{\kappa}, \hskip2mm  Bi  = \frac{ H_{th} d}{\lambda}
\end{eqnarray}
appear.

\noindent  In the following, we discriminate between  two opposite cases:  $Bi \ll  1$ or $1 \ll Bi$.
This  corresponds respectively to an effective  conductance $ H_{th} $ in the gas much smaller or much larger
than the heat conductance $  \lambda/d $  in the fluid.

\section{Basic state and optimal linear perturbations}
\label{linearoptimal}

\begin{figure}
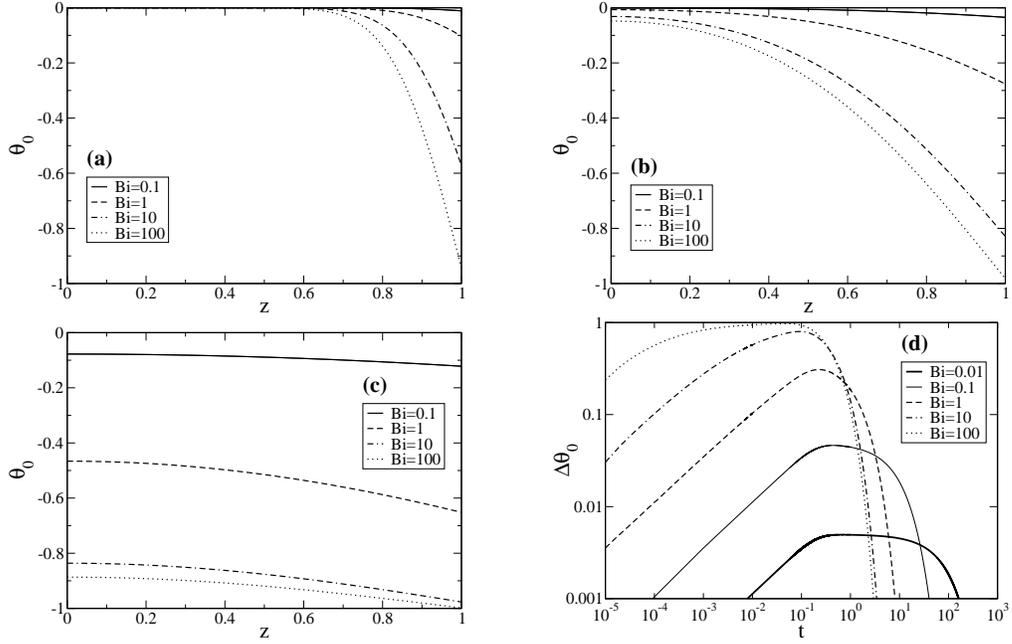

\scriptsize{
\centerline{
\includegraphics[width=0.45\textwidth,clip=]{pics/plot_time1e-2.eps}
\hfill\includegraphics[width=0.45\textwidth,clip=]{pics/plot_time1e-1.eps}
}
\centerline{
\includegraphics[width=0.45\textwidth,clip=]{pics/plot_time1e0.eps}
\hfill\includegraphics[width=0.45\textwidth,clip=]{pics/BaseState.eps}
}
}
\caption{Basic temperature profile  for different Biot numbers.
(a-c)  Temperature profile at   times  $t=10^{-2}$, $t=10^{-1}$ and  $t=1$.
(d) Temperature difference $\Delta {\theta_0}(t)$  as a function of  time $t$.
} \label{basictemperature}
\end{figure}

\noindent We study  the stability of a  purely conductive unsteady
basic state  $\theta_0(z,t) $ which  is  initially  uniform i.e.
$\theta_0(z,t=0) =0 $. This state evolves  since the upper free
surface is cooled down by the latent heat released through
evaporation : The velocity  field  remains always  zero  but  the
unsteady field $\theta_0(z,t)$ satisfies  a pure  heat equation
\begin{eqnarray}
\label{model-heatBAsic}
 \frac{\partial \theta_0}{\partial t} =\frac{ \partial^2  \theta_0}{\partial z^2} ~~\hbox{with }~~\partial_z \theta_0 =0~~~\hbox{at z=0},  \hskip 0.5cm \partial_z \theta_0  +Bi\, \theta_0  +Bi=0~~~\hbox{at z=1}
\end{eqnarray}
The basic  state  only depends on  the Biot number $Bi$. Note that
the term   $Bi \theta_0$   characterizes the heat transfer  in the
gas phase    and   the     term $Bi $ represents the cooling
effect  due to evaporation.   The unsteady temperature field
$\theta_0(z,t)$  is shown in figure~\ref{basictemperature} for
different Biot numbers and    times. A cooled layer develops from
the upper surface whose  thickness $\delta_{0}(t)  \sim \min(\sqrt
t,1)  $   is similar at a given  time for different Biot numbers
and  increases  until it fills the whole layer.  Conversely,  if
$\Delta {\theta_0}(t)$  denotes the characteristic  temperature
difference  within the fluid, the maximum reached over  the whole
time evolution by this quantity  increases with Biot number $Bi$
(figure~\ref{basictemperature}d). Actually,  two different regimes
are observed according to the value of $Bi$ \footnote{Details are
contained  in  Appendix A.}. For small Biot numbers, typically $Bi \lesssim1$,
the  cooled layer  reaches the bottom  while the jump  $\Delta
{\theta_0}(t)$  is less than one. Afterwards,  $\Delta
{\theta_0}(t)$  decreases.    Such   time  evolution  can be
summarized  by the  following scalings
\begin{equation}
\label{CooledLayerBC}
 |\theta_0  | \sim \Delta {\theta_0}    \sim  Bi\sqrt{t} ,~~~~\delta_{0}(t) \sim \sqrt{t} ~~~\hbox{for }~~0 \lesssim \sqrt{t} \lesssim  1,
\end{equation}
  \begin{equation}
\label{CooledLayerBC1}
|\theta_0  | \sim   Bi  ~t,~~~~ \Delta {\theta_0}(t)   \sim  Bi,~~~ \delta_{0} (t) \sim 1  ~~~\hbox{for }~~  1 \lesssim  t \lesssim Bi^{-1},
\end{equation}
For $ Bi^{-1}  \lesssim  t $, the   temperature field  $\theta_0$ relaxes
towards the steady   uniform temperature  $\theta_0 (z,t)= -1$ and
all the energy needed by evaporation is   carried on by convection in the gas phase.

\noindent For large Biot numbers, typically $Bi \gtrsim10$,  the
temperature jump  $\Delta {\theta_0}(t)$   reaches a maximum
before the  cooled layer  reaches the bottom (figure
\ref{basictemperature}a). In that case, the  maximal jump  is
equal to one  and    $|\theta_0(z=1,t) | \sim 1$ at that time.
The  time  evolution  can be summarized  by the  following
scalings :
\begin{equation}
\label{CooledLayerBC3}
|\theta_0  | \sim \Delta {\theta_0}    \sim  Bi\sqrt{t},~~~~\delta_{0}(t)  \sim \sqrt{t}  ~~~\hbox{for }~~0 \lesssim   t \lesssim  Bi^{-2} ,~~~
\end{equation}
\begin{equation}
\label{CooledLayerBC1b}
 \Delta {\theta_0}   \sim 1,~~~ \delta_{0}
\sim \sqrt{t}~~~\hbox{for time}~~  Bi^{-2} \lesssim  {t}  \lesssim  1.
\end{equation}
\noindent
For $ 1  \lesssim  t $,  the temperature decreases in the whole layer thickness
to reach the steady state regime $\theta_0 (z,t)= -1$.


\noindent  In order to  study  the stability of  the unsteady conductive state  we split the   fields into a  basic flow    and
three-dimensional perturbations
\begin{equation}
\label{Decomposition}
\boldsymbol{v} =   \boldsymbol{v}_p(x,y,z,t),~~\theta=\theta_0(z,t) +   \theta_p (x,y,z,t),~~p= p_0(z,t)+p_p(x,y,z,t)
\end{equation}
and   linearize in the   perturbations to get a set of linear equations.
\noindent  Since  this problem has no preferential direction in the $(x,y)$ plane,
the perturbation Fourier modes  in such  directions  decouple in  the linear regime.  Without loss of generality, we  thus consider
a nondimensional  wavenumber $k$ in the $x$ direction
 and no dependence in the $y$ direction  reducing the flow to be  two-dimensional
 \begin{equation}
\label{Fourier}
(\boldsymbol{v}_p,   \theta_p,p_p) =({\hat u}(z,t),0,{\hat w}(z,t),{\hat \theta}(z,t),{\hat p}(z,t))\exp(i kx ).
\end{equation}
\noindent  These infinitesimal perturbations are governed by the linear system
\begin{eqnarray}
\label{Direct1}
\frac{1}{Pr} \frac{ \partial}{\partial t}  {\hat u} + i k {\hat p}
- \left[\frac{ \partial^2}{\partial z^2} - k^2  \right] {\hat u} =0,\\
\label{Direct3}
\frac{1}{Pr} \frac{ \partial}{\partial t} {\hat w} +\frac{ \partial {\hat p}}{\partial z}
- \left[\frac{ \partial^2}{\partial z^2} - k^2  \right]{\hat w}
-Ra  {\hat \theta }  =0,~~~~  ik {\hat u} + \frac{ \partial {\hat w} }{\partial z} =0 ,\\
\label{Direct4}
 \frac{ \partial}{\partial t} {\hat \theta}  + {\hat w}\frac{ \partial {\theta_0 } }{\partial z}
- \left[\frac{ \partial^2}{\partial z^2} - k^2  \right]{\hat \theta }=0, \\
\label{Direct7}
 {\hat w}=0,~~\partial_z {\hat u}  + Ma\, i k {\hat \theta}  =0,~~~~ \partial_z  {\hat \theta} +Bi\,  {\hat \theta} =0,    \hskip3mm\mbox{at $z=1$},\\
\label{Direct6}
{\hat u}= {\hat w}=0, ~~ \frac{\partial {\hat \theta}}{\partial z}=0  \hskip3mm\mbox{at $z=0$}.
\end{eqnarray}

\noindent To quantify the amplification gain at time $t_1$, it is customary to define a norm   which is  generally  based on the kinetic energy of  perturbations.  This  norm can be orthogonally decomposed  in a Fourier basis   so that the individual
contributions of each wavenumber~$k$ can be studied independently. In the present case,
the temperature  field is  playing a major role as well. We thus define   two  different norms
corresponding to  two different situations.   The first one  is based on the kinetic energy of   perturbations
\begin{equation}
E_V (t_1) \equiv   \int ( {\hat u}(z,t_1) {\hat u}^{+}(z,t_1)   + {\hat w}(z,t_1) {\hat w}^{+}(z,t_1)) {\rm d} z
\label{energyphys}
\end{equation}
\noindent where superscript  $+$  denotes  complex conjugation. The integration is
performed over the entire  layer width and   perturbations are obtained after integrating the above  linear system over the time
period $[0,t_1]$.  The second one
\begin{equation}
E_T(t_1) \equiv   \int  {\hat \theta}(z,t_1) {\hat \theta}^{+}(z,t_1)   {\rm d} z
\label{energyphystheta}
\end{equation}
is based on the temperature field and not   the velocity field.

\noindent For finite Prandtl number two extreme cases are
considered, with the initial perturbation concerning either the
velocity field, or  the temperature field.  
In the following we will consider the
amplification factors $E_V (t_1)/E_V (0)$ or $E_T (t_1)/E_T (0)$ to characterize the stability.
Then, when  the initial perturbations only concern the velocity field, 
we only take into account  the amplification  $E_V (t_1)/E_V (0)$   since
$E_T (0)=0$. Conversely, when  the  initial perturbations  only concern the temperature field
we use the amplification factor $E_T (t_1)/E_T (0)$.  
For the infinite Prandtl number case, the velocity perturbations are
not  dynamical quantities since the time derivative of the
velocity drops out from equations (\ref{Direct1})-(\ref{Direct3}).
In other words, the  velocity perturbations  are slaved to the
temperature  perturbations. In this case,  we use only   $E_T
(t_1)/E_T (0)$ and hence the norm  $E_T$.

\noindent  For a given initial disturbance, one   evaluates the
amplification gain  at time $t_1$  by computing $E (t_1)/E (0)$
where $E= E_V$ or  $E= E_T$. It is the purpose of a non-normal
analysis  to   compute the quantity  ${\hat G}(t_1;k;Ma,Ra,Bi,Pr)
\equiv  Max [E(t_1)/E(0)]$ which is the upper bound  for the
energy amplification that  a disturbance of wavenumber  $k$ can
reach at time  $ t_1$. This approach solves a sort of  the finite
time
 stability problem : it investigates  the
transient evolution  and defines  for any time $t_1$,  an optimal
perturbation mode  which actually reaches the upper
bound~(\cite{Schmid2001}). This  mode (i.e. the optimal  initial $z-$ profile) 
is found numerically by solving an optimization problem
(\cite{Farrell96,Anderson99,Luchini00,Schmid2001}). 
The optimum of $E(t_1)$ is determined taking into account
 several  constraints:
(i) the   disturbance energy  at time $t=0$ is equal to  unity; (ii)
the   disturbance  satisfies the linear governing  equation as well as the boundary conditions
during the complete time interval  $[0,t_1]$.  This problem  is best solved with the help
of a Lagrangian formalism and  Lagrangian multipliers which are introduced  to precisely enforce the above
constraints.  In the present case, these multipliers are adjoint fields
$({\tilde u}(z,t),{\tilde w}(z,t),{\tilde \theta}(z,t),{\tilde p}(z,t))$.
Following a  standard derivation, these quantities satisfy
a set of   adjoint equations. It is
\begin{eqnarray}
\label{Adj1}
 \frac{1}{Pr} \frac{ \partial}{\partial \tau}  {\tilde u} - i k {\tilde p}
- \left[\frac{ \partial^2}{\partial z^2} - k^2  \right] {\tilde u} =0,\\
\label{Adj3}
 \frac{1}{Pr} \frac{ \partial}{\partial \tau}  {\tilde w} -   \frac{ \partial  {\tilde p}}{\partial z}
- \left[\frac{ \partial^2}{\partial z^2} - k^2  \right] {\tilde w} + {\tilde \theta}\frac{ \partial {\theta_0 } }{\partial z}
=0, ~~ ~~ ik {\tilde u} + \frac{ \partial{\tilde w} }{\partial z} =0,  \\
\label{Adj4}
  \frac{ \partial}{\partial \tau} {\tilde \theta}
- \left[\frac{ \partial^2}{\partial z^2} - k^2  \right]{\tilde \theta }  -Ra  {\tilde w }  =0,  \\
\label{Adj6}
{\tilde u}= { \tilde w}=0, ~~ \frac{\partial {\tilde \theta}}{\partial z}=0  \hskip3mm\mbox{at $z=0$},  \\
\label{Adj7}
 {\tilde w}=0,~~\partial_z {\tilde  u}   =0,~~~~ \partial_z  {\tilde \theta} +Bi\,  {\tilde \theta} + Ma\, \partial_z  {\tilde w} =0    \hskip3mm\mbox{at $z=1$},
\end{eqnarray}
\noindent  in which  $\tau \equiv -t$.  These  adjoint  equations   have to be solved backwards in time.   Let us denote by the symbol
 $\boldsymbol{q}$   the   vector field   $({u},{w}, {\theta},{p})$. One obtains the optimal perturbation for time $t_1$ by an iterative scheme
  which propagates a given initial condition forward in time using the direct problem (here denoted by  $F_j(\boldsymbol{q})=0,~~ j=1 \ldots 4$), the result of
which serves as an ``initial'' condition for the backward propagation
by the adjoint equations (here denoted by  ${\tilde F}_j(\tilde{\boldsymbol{q}}(t))=0, j=1 \ldots 4$). 
 More
specifically \footnote{Details are contained  in  Appendix B.}, a relation between the adjoint  $\boldsymbol{\tilde{q}}(z,t_1) $
and $\boldsymbol{{q}}(z,t_1) $ is imposed.  After one forward-backward integration,  quantity   $\boldsymbol{\tilde{q}}(z,0) $ is obtained
 and a relation between $\boldsymbol{\tilde{q}}(z,0)$ and $\boldsymbol{q}(z,0) $ is also imposed.
An updated initial condition for the next iterative step is  then available.
This process should be self-consistent :  one   uses an iteration procedure which is schematically
illustrated by a diagram
\begin{equation}
\label{DIAGRAM}
\begin{array}{ccc}
\boldsymbol{q}(z,0) &\stackrel{F_j\left(\boldsymbol{q}\right)=0}{\longrightarrow}
&\boldsymbol{q}(z,t_1)\\
\uparrow&&\downarrow\\
\tilde{\boldsymbol{q}}(z,0) &\stackrel{\tilde{F}_j\left(\tilde{\boldsymbol{q}}\right)=0}{\longleftarrow}
&\tilde{\boldsymbol{q}}(z,t_1)
\end{array}
\end{equation}
\noindent  Convergence is reached when the initial
condition for the forward problem does not change appreciably  -- up to a
normalization constant --  by an appropriately chosen criterion from one iterative step to the
next.  The converged mode is precisely the  initial optimal perturbation for time $t_1$.
The maximum energy amplification is    computed by
propagating the converged initial condition forward in time and by
forming the ratio of the disturbance energy at the end of the time
interval to the energy at the beginning.
The direct and adjoint equations   have been discretized using a
pseudospectral scheme based on   Chebyshev polynomials and a streamfunction-based formulation
to account for incompressibility\footnote{Details are contained  in  Appendix  C.}.


\section{Numerical Results}
\label{results}

\subsection{Quantities provided by the non-modal analysis}
\label{resultquantity}

\begin{figure}
\scriptsize{
\centerline{
\includegraphics[width=0.45\textwidth,clip=]{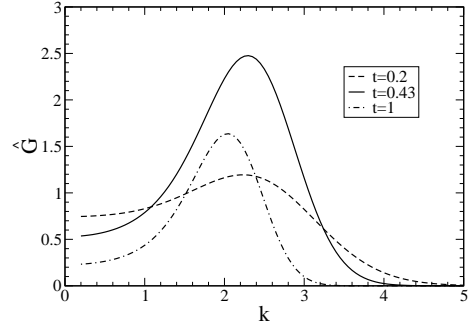}
}}
\caption{ The maximum energy amplification ${\hat G}(t;k;Ma,Ra,Bi,Pr) $ as a function of  wavenumber $k$  for three different  times  $t=0.2$,  $t=0.43$  and $t=1.$
Parameters $Ma=300$, $Ra=0$, $Bi=1$, $Pr= \infty$.
}
\label{DispersionRelation}
\end{figure}

\begin{figure}
\scriptsize{
\centerline{
\includegraphics[width=0.45\textwidth,clip=]{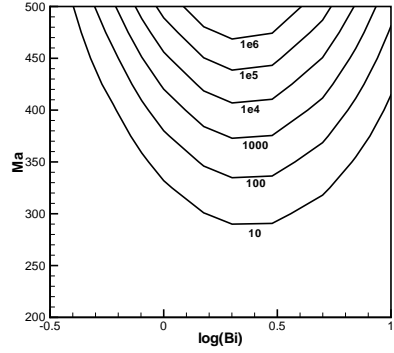}}}
\caption{ Isolines of the  maximum amplification $  G_{max}
(Ma,Ra,Bi,Pr)  $ in the plane  $(Bi,Ma)$
Parameters   $Ra=0$, $Pr= \infty$.  The values of the isolines are written on the figure.
}
\label{MaxiAmpli}
\end{figure}

\noindent For the unsteady basic state $\theta_0(z,t)$,   a given
set  $(Ma,Ra,Bi,Pr)$, and a given type of initial perturbation
(temperature or velocity), the non-modal analysis determines at
 a given time  $t_1$,   the maximum energy amplification ${\hat G}(t_1;k;Ma,Ra,Bi,Pr) $
over all possible  perturbations  of  wavenumber $k$ (see figure
\ref{DispersionRelation}).  In a way,  the value $\ln{\hat G/t_1}$
is analogous to  a  growth rate for classical stability analysis.
More generally, it  appears possible to extend the usual concepts
of  classical stability analysis to unsteady flows.  For instance,
${\hat G}(t_1;k;Ma,Ra,Bi,Pr)$ can be maximized over wavenumber $k$
and time $t_1$  providing  the global maximum amplification
$G_{max}(Ma,Ra,Bi,Pr)$ (see figure \ref{MaxiAmpli}). This value is
reached at time $ t=t_{opt}(Ma,Ra,Bi,Pr)$,  for an optimal
wavenumber  denoted by $k_{opt}(Ma,Ra,Bi,Pr)$ and   for a specific
perturbation  structure in $z$.   These latter two  quantities
play the role of the most amplified wavenumber   and   of  the
most amplified  mode  for the standard stability analysis. One
also obtains a  "stability" diagram,  by determining  the region
of the space $(Bi, Pr, Ma, Ra)$ where the amplification gets above
a threshold $G_{thres}$. It is a way of   separating  the region
where amplification  or  attenuation occurs. The  value of  $G_{thres}$, for instance $G_{thres}=1$,
 is somewhat arbitrary : as already said in the introduction, an unsteady problem
is indeed characterized  more by a transition domain  than  a well defined
threshold.   It is demonstrated below that choosing
$G_{thres}=1$ or $G_{thres}=100$  does not result in major
differences, so that the transition region is thin compared with the
absolute values of the Marangoni and Rayleigh numbers. For fixed $Ra$
and $Pr$,  one can determine the curve $Ma_c(Bi,Ra,Pr)$ such that
if $Ma<Ma_c$ (resp. $Ma
>Ma_c$) then $  G_{max} <G_{thres}$ (resp. $  G_{max} >
G_{thres}$). Similarly one may define for fixed   $Ma$ and  $Pr$,
 the  curve   $Ra_c(Bi,Ma,Pr)$.
These curves  play a role very much similar to  marginal stability curves.
Each point of the  critical curve is associated to a critical wavenumber
 $k_c \equiv k_{opt} $ and critical optimal time $t_c \equiv   t_{opt}$.
Note that until this point, most of this procedure can be extended
to other unsteady problems in a straightforward way.  A comparison
between these critical curves and the experimental diagram that
separates the domains where convection is observed or not observed,
is made in section \ref{Experiment}.


\subsection{ Infinite Prandtl number:  the pure Marangoni case  $ Ma \neq 0$ and $Ra=0$.}
\label{resultPrandtlNinfiniteMa}

\noindent For infinite Prandtl number, the velocity is slaved to the temperature field. As a consequence, only perturbations in temperature field   are pertinent.   In the plane $(Bi,Ma)$,   the  critical curve  $Ma_c(Bi)$,  critical wavenumber $k_c(Bi)$
and critical optimal time $t_c(Bi)$ are presented for the  pure Marangoni case and  two thresholds $G_{thres}=1$ and
 $G_{thres}= 100$ (figure \ref{CriticalMarangoniPrINF}).
 The critical Marangoni  number $Ma_c(Bi)$    slightly depends on the value of the threshold and  seems to be consistent, within the numerical uncertainties, to the following laws (here  $G_{thres}=1$)
\begin{eqnarray}
Ma_c(Bi) \simeq 83/ Bi   ~~~~\mbox{for}~~~Bi \ll 1  \hskip5mm\mbox{and} ~~~~Ma_c(Bi) \simeq  15  Bi  ~~~\mbox{for}~~1 \ll Bi
\label{MacBi}
\end{eqnarray}
  \noindent Moreover the  wavenumber  $k_c(Bi)$ (figure \ref{CriticalMarangoniPrINF}b) is an increasing function of the Biot number while
optimal time  $t_c(Bi)$ is a decreasing function of the same parameter.

\noindent Using heuristic arguments \footnote{Details are contained  in  Appendix D.},   the following   scaling laws can be deduced : \\
\noindent For $Bi  \ll   1 $  the critical  conditions for instability  can be expressed  as
\begin{equation}
\label{OnsetGENERAL1sup}
Ma_c \sim    1/Bi  ~~ \hbox{with}~~~\sqrt{Bi} \lesssim   k_{c}   \lesssim   1 ~~ \hbox{and}~~~ 1  \lesssim t_{c}  \lesssim  1/Bi.
 \end{equation}
\noindent For $  1 \ll Bi$, the critical  conditions become
\begin{equation}
\label{OnsetGENERAL2sup}
Ma_c \sim   Bi  ~~  \hbox{with}~~~1 \lesssim   k_{c}   \lesssim   Bi ~~ \hbox{and}~~~      t_{c}   \sim   k_c^{-2}
 \end{equation}

\begin{figure}
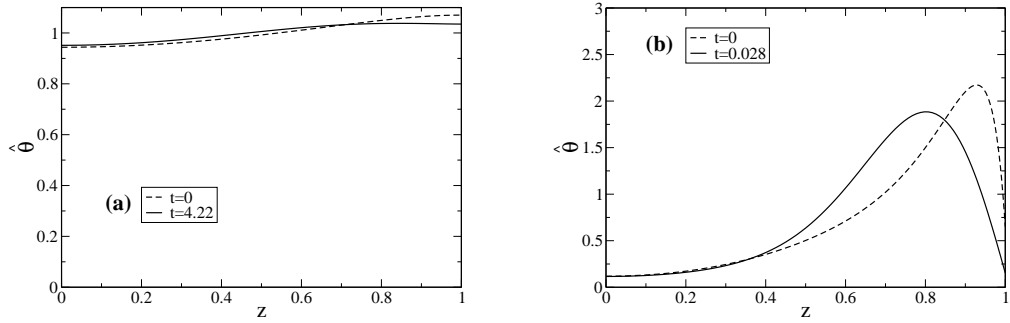

\scriptsize{
\centerline{
\includegraphics[width=0.45\textwidth,clip=]{pics/thetaopt-mapurebi0.01.eps}
\hfill\includegraphics[width=0.45\textwidth,clip=]{pics/thetaopt-mapurebi100.eps}
}}
\caption{The optimal   temperature perturbation  ${\hat \theta}(z,t)$   at  time $t=0$ (dashed) and $ t=t_{c}$ (solid)   for
$Ra=0 $, $Pr= \infty$, $G_{thres}=1$ :  (a) $Bi=0.01$, $Ma=Ma_{c}=8685$ and $k=k_c=0.74$,   (b) $Bi=100$, $Ma=Ma_c=1658$ and
$k=k_c=4.46$.
 }
\label{StructureMarangoniPrINF}
\end{figure}

\noindent   The spatial structure in $z$  of the   optimal perturbation at $k=k _c$ and $Ma=Ma_c$    is shown on
 figure~\ref{StructureMarangoniPrINF} for two  Biot
 numbers and two different times~: time $t=0$ and time $t=t_c$
when the perturbation reaches its maximum amplification. The spatial structure
of this optimal perturbation is shown to change slightly during
the  time evolution. In this respect, this optimal mode does not
differ much from  the classical most amplified mode of steady
problems.


\subsection{ Infinite Prandtl number : the pure Rayleigh case   $Ra \neq 0$ and $ Ma = 0$.}
\label{resultPrandtlNinfinite}

\noindent  In the plane $(Bi,Ra)$,   the  critical curves $Ra_c(Bi)$,   $k_c(Bi)$
and  $t_c(Bi)$ are presented  for  two thresholds $G_{thres}=1$ and
 $G_{thres}= 100$ (figure~\ref{CriticalRayleighPrINF}). The critical Rayleigh
  $Ra_c(Bi)$  slightly depends on the value of the threshold and  seem to be consistent,
  within  numerical uncertainties,  to the following laws (here  $G_{thres}=1$)~:
 \begin{eqnarray}
Ra_c(Bi) \simeq 600/ Bi   ~~~~\mbox{for}~~Bi \ll 1 \hskip5mm\mbox{and} ~~~~~~Ra_c(Bi) \simeq  960~~~\mbox{for}~~~1 \ll Bi
\label{RacBi}
\end{eqnarray}
 \noindent Moreover the  wavenumber  $k_c(Bi)$  (resp.  time $t_c(Bi)$) is an increasing (resp. decreasing)  function  of
 the Biot number for small Biot  and reaches a plateau for larger Biot number.

 \noindent One can deduce using heuristic arguments \footnote{Details are contained  in  Appendix D.},   the following  scaling laws:
 \begin{equation}
\label{OnsetGENERAL1RBsup}
Ra_c \sim    1/Bi  ~~ \hbox{with}~~~\sqrt{Bi} \lesssim   k_{c}   \lesssim   1 ~~ \hbox{and}~~~ 1  \lesssim t_{c} \lesssim  1/Bi, ~~\hbox{for}~~Bi  \ll   1
 \end{equation}
\begin{equation}
\label{OnsetGENERAL2RBsup}
Ra_c \sim   1  ~~  \hbox{with}~~~ k_{c} \sim 1  ~~ \hbox{and}~~~      t_{c}   \sim   1,~~\hbox{for}~~1  \ll   Bi.
 \end{equation}


%


\subsection{Comparison with classical steady results and with transient frozen-time method}

\noindent  It is worth  comparing  the  results presented  here
with the well-known results obtained  by  \cite{Pearson:1958} 
\footnote{In this case, we computed some additional results to cover a larger range of Biot 
numbers than the one  given by Pearson in his article. }
(resp. \cite{Sparrow63}) in the framework of  the steady Marangoni
(resp. Rayleigh) problem. This comparison is pertinent since the
boundary conditions at the top and the bottom of the layer
(equations \ref{Direct7}-\ref{Direct6}) are
similar  in the present work and in these classical analyses.
However, the Marangoni and Rayleigh numbers defined by these
authors are based on the  steady temperature difference $\Delta
T_0$ between the top and the bottom of the layer. 
This steady
temperature difference is missing in the transient problem under
study.  It is then not possible to make a direct comparison between 
the thresholds values obtained in our paper
and  the previous ones from Pearson's or Sparrow's publications, 
and a preliminary transformation is needed. 
Indeed,   at each time $t$, one might define the equivalent
temperature difference between the top and the bottom of the layer
i.e. $\Delta \theta_0(t) \Delta T_{st}$. Using such a temperature
difference, it is then  easy to define a time-dependent Marangoni $\bar
Ma (t)= \Delta \theta_0(t) Ma$  or a time-dependent Rayleigh
numbers $\bar Ra(t) = \Delta \theta_0(t) Ra$.

\noindent Let us compute  the   maximum  of   $\Delta \theta_0(t)$ obtained  during the   time evolution 
for the Marangoni  (resp. Rayleigh) case. This maximum is
reached at  time $ t_c$ and leads to   a new  Marangoni  number $\bar
Ma(t_c) $ (resp. Rayleigh number $\bar Ra (t_c) $)  
which can be compared to the critical  values $\bar Ma_{Steady}$
(resp. $\bar Ra_{Steady}$)   obtained  by  \cite{Pearson:1958}
(resp. \cite{Sparrow63})  from the  steady case.
\noindent  With the scalings  used here, the  critical Marangoni $
Ma_c$ predicted from these steady results  reads
 \begin{equation}
\label{Anciens} Ma_c \simeq \frac{\bar Ma_{Steady}}{\Delta
\theta_0( t_c)}   ~ \hbox{and} ~  Ra_c \simeq \frac{\bar
Ra_{Steady}}{\Delta \theta_0( t_c)}
\end{equation}
\noindent Such estimations have been plotted on figure
\ref{CriticalMarangoniPrINF} and \ref{CriticalRayleighPrINF}. They
 are found to be   close to our results for  $G_{thres}= 1$.
 The same comment
applies for the critical wavenumbers except for 
the B\'enard-Marangoni  case at  high Biot numbers.
For  critical times,  however, the  steady-state approximation differs from our results.

\noindent When using the results by \cite{Pearson:1958} or \cite{Sparrow63}, we
are clearly  using the normal mode results obtained for a linear
temperature field in $z$ on the whole thickness, which is a rough approximation 
especially at high Biot number. We can go even  further and
compare the non-normal mode results   within  the
frozen-time approximation. In this  latter approximation, 
a stability  analysis in terms of   
normal modes  is   performed   at the each time $t$. 
 The "steady" base flow is assumed to be
the temperature field $ \theta_0(z,t)$ computed at  this specific time $t$.
When the flow  is stable  within this frozen-time approximation for each time $t$, it will 
be assumed stable. When  the  control parameter reaches  a  critical 
value (here  $ Ma_c$ or $ Ra_c$),  there exists a unique  critical time $t_c$  for which the frozen time 
state $ \theta_0(z,t_c)$ possesses a marginal eigenvector with a critical wavenumber.
In the present problem we have determined the critical parameters for neutral
conditions in the frozen-time case by  the Lapack routine GGEV for generalized eigenvalue problems
in combination with  a  Chebyshev collocation
method. As can be seen in figures \ref{CriticalMarangoniPrINF} and
\ref{CriticalRayleighPrINF}, results for the thresholds and the
critical wavenumbers using the non-normal approach ($G_{thres}= 1$) and
the  frozen-time approximation  are close. However the prediction of critical times
$t_c$ still differs~\footnote{In the frozen-time approximation,  $t_c$ 
corresponds to  a time   when  the quasi-static growth rate  $\sigma(t)$ becomes zero. 
This critical time is of a different nature than the critical time for the non-normal 
 approach. The latter characterizes the perturbation evolution over the time interval from
 $t=0$ up to $t_c$. 
In  zeroth order WKB theory,  the critical time $t_c$ for the non-normal approach
would be determined by $ \int^{t_c}_{0} \sigma(t) dt = \ln(G_{thres})$. 
Suitable modifications of the frozen-time analysis based on this observation should
therefore lead to closer agreement regarding the critical times. We do not pursue this
issue further in the present work.}.
The   frozen-time approximation apparently provides a bound of order
unity  for $t_c$ because the available temperature difference $\Delta \theta_0( t)$ 
attains its maximum as soon as the thermal boundary layer of 
the basic temperature distribution has reached the bottom
of the layer. For small $Bi$,  $\Delta \theta_0( t)$  remains  fairly
constant for larger times, and the instability can develop on 
this quasi-steady background over  fairly long times. This observation
can explain the apparently unbounded growth of $t_c$ with $Bi$ 
for $Bi\to 0$ in the non-normal analysis.

\subsection{ Results for  finite Prandtl numbers}
\label{resultPrandtlNfinite}

\begin{figure}
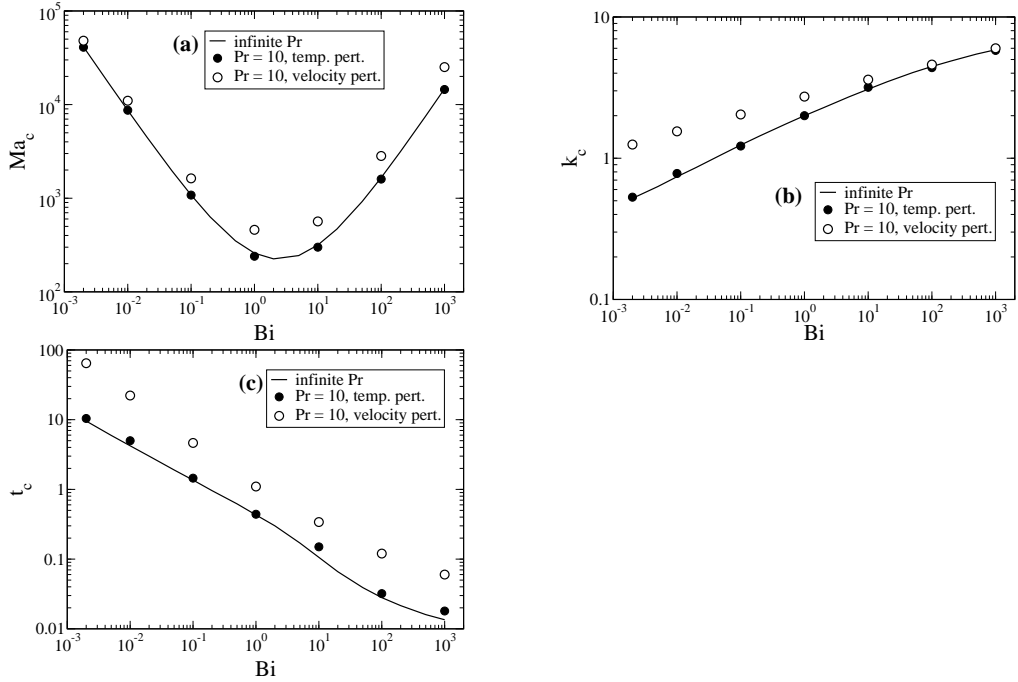

\scriptsize{
\centerline{
\includegraphics[width=0.45\textwidth,clip=]{pics/MacPrFini.eps}
\hfill\includegraphics[width=0.45\textwidth,clip=]{pics/MaKcPrFini.eps}}
}
\centerline{
\includegraphics[width=0.45\textwidth,clip=]{pics/MaTcPrFini.eps}
\hfill
}
\caption{ Infinite  Prandtl ( $Pr= \infty$) or finite Prandtl  ($Pr=10$) number cases. Temperature or velocity initial perturbation results  with threshold  $G_{thres}=1$.  (a) Critical Marangoni $Ma_c(Bi)$, (b) Critical wavenumber $k_c(Bi)$, (c) Critical time $t_c(Bi)$.
}
\label{MarangoniFinitePr}
\end{figure}

\noindent In this  section, we focus on  finite Prandtl numbers and more specifically on the role of
initial perturbations on the transition zone estimation. In this case, the  velocity field  is not slaved
to the temperature field  so that  perturbations in velocity can be
considered  as well as perturbations  in  temperature. 
  We only  discuss the  curves for the pure Marangoni case
 ($Ra=0$)~(Figure \ref{MarangoniFinitePr}) but similar results apply to   the pure Rayleigh case ($Ma=0$).\\

\noindent  Since  the frozen time approximation seems to be valid for this unsteady problem and 
an exchange of stability in a normal mode is not affected by the  Prandtl number, 
the effect of  this latter number should not be significant. 
Indeed, for  temperature perturbations,   
the critical Marangoni $Ma_c(Bi,Pr)$  is not  affected  
when infinite $Pr$ number case is compared to $Pr=10$. 
For the optimal time $t_c (Bi,Pr)$ and   wavenumber $k_c (Bi,Pr)$, the same
conclusions apply. For  velocity  perturbations,  the curves
differ but not in a drastic way taking into account that
perturbations are of a completely different nature  compared with
the infinite-Prandtl-number problem. Perturbing the temperature is
more efficient than perturbing the velocity. Indeed critical
Marangoni  numbers are smaller for temperature perturbations, but
the difference is of the same order as the one obtained by
changing the threshold value from 1 to 100. Actually, the ``blurriness'' 
of this transient problem induced by the choice of the threshold
values
 and perturbation types
 is not very broad and does not   modify the order of magnitude of the  critical numbers if one excepts the time $t_c$.
Finally, let us note that non-linear direct simulations have also
been made to solve this problem(\cite{touazi09}), showing very good
agreement with the linear results presented here.



  \section{ Comparison with experiments}
  \label{Experiment}

\noindent   The  present section is devoted to the  comparison between  results  obtained from the optimal mode calculations and  from an experimental work  described in \cite{tous08}, where transient   Rayleigh-B\'enard-Marangoni  convection is  generated by  drying   a polymer solution of PolyIsobutylene-Toluene  at ambient temperature.
In the experiments,  buoyancy and Marangoni effects are  equally present.  It is the
evaporation  of the solvent  i.e. toluene which cools the upper surface by latent heat. During the experiments,  the following
parameters are kept constant:

\begin{eqnarray}
\kappa=  0.97 \times 10^{-7}~~m^2~s^{-1},~~\lambda = 0.142~~W~K^{-1}~m^{-1},~~\rho = 865~~ kg~m^{-3},~\nonumber{}\\
\alpha =  1.07 \times 10^{-3}~~K^{-1},~~\sigma=28 \times 10^{-3}~~N~m^{-1},~~\gamma  =   1.19 \times  10^{-4}~~N~K^{-1}~m^{-1},~~~ \\
L=3.96 \times 10^5 J~kg^{-1},~~~H_{th} = 28~  W~K^{-1}~m^{-2},~~~ \Delta T_{st}=4.8~K.~~~  \nonumber{}
\end{eqnarray}
Different  thicknesses $d$  and dynamic  viscosities   $\mu$  are considered.  $d$
is varied from $0.3~mm$ to $23.5~mm$ while  dynamic viscosity  $\mu$  is set to a  value in the range  $[0.55~mPa~s~,~2100~mPa~s]$
by monitoring the initial polymer concentration. These data allow us to estimate the crispation, the Galileo and the Peclet numbers for each experiment. We get the following bounds for these numbers:
\begin{equation}
10^{-7} \leq Cr \leq  10^{-3}, ~~~ 2 \times 10^2 \leq Ga \leq 3 \times 10^8,~~~10^{-3} \leq Pe \leq 0.1
\end{equation}

\begin{figure}
\scriptsize{
\centerline{
\includegraphics[width=0.45\textwidth,clip=]{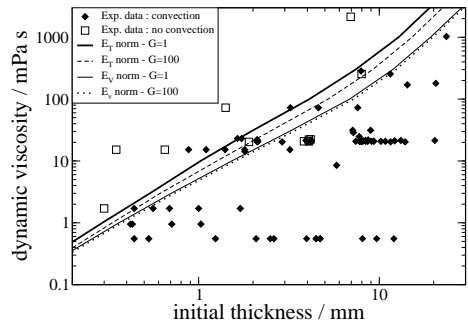}}
}
\caption{Comparison of theoretical and experimental results in the plane $(d ,\mu)$ where $d$ stands for the layer  thickness and $\mu  $
the  dynamic  viscosity.  The various curves displayed in solid lines correspond to two different thresholds  $G_{thres}=1$, $G_{thres}=100$  and  two different perturbation
 types (velocity and temperature).
Experimental data are displayed by symbols.
 }
\label{ExpTheor}
\end{figure}

\noindent As a consequence, the assumption of planar free surface and constant thickness layer are justified. The other relevant non-dimensional numbers vary in the following range :
\begin{equation}
0.06 \leq Bi \leq 5 ; ~~ 6.6 \leq Pr \leq 2.5 \times 10^4;~~20 \leq Ma \leq 1.2\times 10^5 ;~~1.3 \leq Ra \leq 1.4\times 10^6
\end{equation}
 \noindent  The comparison  is  displayed  using  the  critical   dynamic viscosity  $\mu_c(d) $    as a function of thickness $d$ (figure \ref{ExpTheor}).   In the   plane  $(d,\mu)$,  each point   corresponds  to a given parameter set   $(Ra,Ma, Pr, Bi )$.   The four  critical curves correspond to  two thresholds ($G_{thres}=1$  and
  $G_{thres}= 100$) and two  types of initial pertubation (temperature and velocity).
The theoretical critical curves  divide the  experimental points corresponding to regions
of convection or pure conduction  in a satisfactory manner. The temperature perturbation critical curve
is above the velocity one. 
 This analysis in the thickness/viscosity plane shows again  that the bandwith of uncertainty due to the choice of threshold $G_{thres} $ and perturbation
 types is not very broad and does not   modify the order of magnitude of the  critical thickness.


\section{Conclusion.}

\noindent  This paper  presents a novel 
linear  stability analysis of an unsteady base state
within the general conceptual framework of amplification theory.
The non-normal approach is used, which possesses the  advantage over 
more classical  methods  to solve  the transient problem  in a 
mathematically rigorous way. In turn, this  allows one to test  other 
approximations. Here we have applied this
approach  for the first time  to characterize the stability of a transient 
Rayleigh-B\'enard-Marangoni problem in an horizontal  fluid layer 
  suddenly cooled from above.
It provides the
upper limit of the energy amplification that a disturbance of
wavenumber $k$ can reach at time $t$. This quantity reaches a
maximum   at time $t=t_{opt}(Ma,Ra,Bi,Pr)$,  for  a specific
optimal wavenumber $k_{opt}(Ma,Ra,Bi,Pr)$ and    a specific
perturbation structure in $z$.   These latter two quantities
play the role of the most amplified wavenumber   and  most
amplified  mode for the standard steady analysis.   

\noindent  A "stability" diagram  in   the   space $(Bi,
Pr, Ma, Ra)$ has been determined 
for the pure Marangoni and the pure Rayleigh
problem by the non-normal approach.
Note that the marginal conditions used to determine the stability curve
was obtained by setting the optimal amplification equal to 1 or 100.
The critical time and critical wawenumber were evaluated for this marginal conditions.
Critical Marangoni and Rayleigh numbers exhibit a
strong dependency on the Biot number and a weak sensitivity to
Prandtl number variations  in the range $Pr \geq 1$. Scaling
exponents for critical Rayleigh or critical Marangoni versus Biot
numbers  have been found  numerically and confirmed  by scaling
analysis  in the limit of very small and very large Biot numbers.
Comparison of the non-normal approach with the frozen-time
approximation (a classical 
quasi-static approach) shows similar results
for the critical Marangoni or Rayleigh numbers and the critical wave numbers.
Moreover, the ``blurriness'' inherent in any
transient problem  was analyzed as a function of the amplification
threshold values
 and perturbation types. It  has been shown that the transition region
 is thin compared to the large domain of Rayleigh or Marangoni
 numbers covered by the analysis.

\noindent Finally, a comparison with experimental results has been performed, where convection is
induced by solvent evaporation during the drying of polymer
solution. A good agreement was indeed found  between the present theoretical study and experimental observations.
The method was developed in this paper for the cooling of a fluid induced by solvent
evaporation but could easily be extended to other transient
problems.

\acknowledgments{ TB and MR acknowledge financial support
from the Deutsche Forschungsgemeinschaft in the framework of the
Emmy--Noether Program (grant Bo 1668/2).}

\appendix


\section{  Basic State}
\label{sec:BasicState}
 
\noindent  We obtain here  the scaling laws given in section 3 which  provide  the evolution of a purely conductive basic state. 
It is  recalled that the basic temperature field is initialy uniform i.e. $\theta_0 (z,t=0)=0$.
 and satisfies 
a heat equation

\begin{eqnarray}
\label{model-heatBAsic-app}
 \frac{\partial \theta_0}{\partial t}    &=&  \frac{ \partial^2  \theta_0}{\partial z^2} 
\end{eqnarray}
with the boundary conditions 
\begin{equation}
\label{model-heatBAsicBC}
 \partial_z \theta_0 =0~~~\hbox{at z=0},  \hskip 2cm \partial_z \theta_0  +Bi\, \theta_0  +Bi=0~~~\hbox{at z=1}.
\end{equation}

\noindent Due to the evaporation,  a cooled layer of characteristic thickness $\delta (t)$ develops from the upper surface. 
Equation (\ref{model-heatBAsic-app}) provides the standard estimate
\begin{equation}
\label{CooledLayer}
	\delta_{0}(t)  \sim \min(\sqrt t,1).
\end{equation}
\noindent  In the following, we  determine the scaling laws for  the two  extreme cases $Bi\ll 1$ and $1 \ll Bi$.

\subsubsection*{A) Case  $Bi \ll  1$  }

\noindent Since $\theta_0 (z,t) $   is  initially zero, it remains small during a period of time  $t \lesssim  \tau_{1} $~(the time $\tau_{1}$  is determined below and  shown  to be much larger than $1$).
Diffusion in the liquid  and evaporation terms thus dominate in the free-surface boundary condition (\ref{model-heatBAsicBC}). 
 Such a  balance can be expressed  in terms of order of magnitude as follows
\begin{equation}
\label{CooledLayerBC0-app}
\frac{\Delta {\theta_0}(t) }{\delta_{0}(t)} \sim  Bi 
\end{equation}
One thus   obtains using equation (\ref{CooledLayer})
\begin{equation}
\label{CooledLayerBC-app}
\Delta {\theta_0}    \sim  Bi\sqrt{t} ,~~~~\delta_{0}(t) \sim \sqrt{t} ~~~\hbox{for }~~0 \lesssim \sqrt{t} \lesssim  1,
\end{equation}
\noindent  During  this   time interval,  the  cooling layer has not reached the bottom at $z=0$ so that  temperature field  $\theta_0$  reads   in terms  of orders of magnitude 
\begin{equation}
\label{CooledLayerBC22}
 |\theta_0  | \sim  \Delta {\theta_0}    \sim  Bi\sqrt{t} ~~~\hbox{for }~~0 \lesssim \sqrt{t} \lesssim  1.
 \end{equation}
\noindent Thereafter the  thickness remains  constant $\delta_{0}(t) \sim 1$  and   the following condition holds 
\begin{equation}
\label{CooledLayerBC1-app}
 \Delta {\theta_0}(t)   \sim  Bi,~~~ \delta_{0} (t) \sim 1  ~~~\hbox{for }~~  1 \lesssim  t \lesssim  \tau_{1}.
\end{equation}
During  this latter time interval,  the   heat equation (\ref{model-heatBAsic-app})  can be used with scaling (\ref{CooledLayerBC1-app}) to get the temperature field  $\theta_0$    in terms  of orders of magnitude 
\begin{equation}
\label{CooledLayerBC2}
 |\theta_0  | \sim   Bi  ~t, ~~~\hbox{for }~~  1 \lesssim  t \lesssim  \tau_{1}.
\end{equation}
\noindent  These equations are valid  if evaporation dominates  heat transfer in the gas phase. This requires that  $|\theta_0 | \ll 1$ 
and determines   the   value    $\tau_{1} \sim Bi^{-1}$.  For $ \tau_{1} \lesssim  t $, the   temperature field  $\theta_0$ relaxes 
towards the steady   uniform temperature equal to $\theta_0 (z,t)= -1$.  Since the temperature gradient is equal to zero in that regime, 
all the energy due to evaporation  is transfered by convection in the gas phase.

\subsubsection*{B) Case $   1  \ll Bi $}

\noindent For small times,   the condition $ |\theta_0 |  \ll  1$  holds and  an  analysis similar to the one performed for the case $Bi  \ll  1$ is valid leading  to 
\begin{equation}
\label{CooledLayerBC3-app}
 \Delta {\theta_0}    \sim  Bi\sqrt{t},~~~~\delta_{0}(t)  \sim \sqrt{t}  ~~~\hbox{for }~~0 \lesssim   t \lesssim {\tau_{2}} ,~~~\hbox{where}~~~{\tau_{2}} \sim Bi^{-2}.
\end{equation}
The value of ${\tau_{2}}$ is obtained  by determining the time when    $\Delta {\theta_0} \sim 1$. At that time,  the heat flux in the gas phase  becomes of the 
same order of the evaporation.    During  this   time interval,  the  cooling layer has not reached the bottom at $z=0$ so that  temperature field  $\theta_0$  reads   in terms  of orders of magnitude 
\begin{equation}
\label{CooledLayerBC23}
 |\theta_0  | \sim  \Delta {\theta_0}    \sim  Bi\sqrt{t} ~~~\hbox{for }~~0 \lesssim {t} \lesssim Bi^{-2}.
 \end{equation}
For $ Bi^{-2} \lesssim t$, a new regime  begins  where  the surface temperature remains constant 
$|\theta_0 (z=1,t) |  \sim 1$    while the cooled layer thickness keeps increasing
\begin{equation}
\label{CooledLayerBC1b-app}
 \Delta {\theta_0}   \sim 1,~~~ \delta_{0} 
\sim \sqrt{t}~~~\hbox{for time}~~  Bi^{-2} \lesssim  {t}  \lesssim  1.
\end{equation}

\noindent    This regime ends  when $\delta_{0} 
\sim 1$ at time $t \sim 1$. Thereafter   the temperature decreases in the whole layer thickness 
to reach the steady state regime $\theta_0 (z,t)= -1$.

\noindent The case $Bi \sim 1$ is the limiting case of the two previous ones. The equation (\ref{CooledLayerBC-app}) is valid until $t \sim 1$, when the cooled layer thickness and the 
surface temperature  both  reach their extremum. Thereafter the temperature decreases in the whole layer thickness, to reach the steady state.


\section{Obtaining the Adjoint Equations}
\label{sec:annexA}

\noindent Let us denote by $q_j(z,t),j=1, \ldots, 4$,  the components of the  
vector field 
$$
({\hat u}(z,t),{\hat w}(z,t),{\hat \theta}(z,t),{\hat p}(z,t)).
$$ 
To find the maximum amplification at a given time $t_1$,
we maximize the  perturbation norm   $E(\boldsymbol{q}(t_1))$ 
\begin{equation}
E(\boldsymbol{q}(t_1)) \equiv \sum^{3}_{j=1} C_j \int q_j(z,t_1) {q}^{+}_j(z,t_1) {\rm d} z
\end{equation}
at time $t_1$ with respect to the set of all possible initial perturbations $\boldsymbol{q}(0)$ such that  $E(\boldsymbol{q}(0))=~1$. It is recalled that the   integration is performed over
the entire layer depth and the superscript $+$
denotes complex conjugation. Coefficients $C_j$ are  weight coefficients chosen to put emphasis on temperature or velocity acoording 
to the case  considered. To analyse the initial perturbation in velocity,  the  kinetic energy norm $E_V$ is used and  one takes  $C_1=C_2=1$ and $C_3=0$. 
To analyse the initial perturbation in temperature,   the temperature norm $E_T$ is used and   $C_1=C_2=0$ and $C_3=1$.

\noindent The variation $\delta E(\boldsymbol{q}(t_1))$ with respect to a variation $\delta \boldsymbol{q}(0)$ of the initial perturbation
is to be  evaluated. This computation cannot be performed
in a straightforward manner since the energy $E(\boldsymbol{q}(t_1))$ can be  explicitly
written in terms of $\boldsymbol{q}(t_1)$ but only implicitely in terms of $\boldsymbol{q}(0)$. It is known 
{\it via} several constraints: normalization of $\boldsymbol{q}(0)$ and  time integration over the  interval $[0,t_1]$ of   
 equations (3.8)--(3.10). These dynamic equations relating  $\boldsymbol{q}(0)$ to $\boldsymbol{q}(t_1)$,  are  formally written here as $F_j(\boldsymbol{q})=0,~~ j=1 \ldots 4$.  This optimization with constraints necessitates
the introduction of Lagrangian multipliers, the so-called adjoint fields
$\tilde{\boldsymbol{q}}(t) \equiv ({\tilde u}(z,t),{\tilde w}(z,t),{\tilde \theta}(z,t), {\tilde p}(z,t))$.

\noindent More specifically,  a Lagrangian function $L$ is  defined, which depends
on direct ${\boldsymbol{q}}(t)$ and adjoint $\tilde{\boldsymbol{q}}(t)$ variables
over the interval $ [0,t_1]$, and a normalization scalar $s_0$:
\begin{eqnarray}
\label{Lagrangian}
L(\boldsymbol{q},\tilde{\boldsymbol{q}},s_0,t_1)=  E(\boldsymbol{q}(t_1)) - s_0(E(\boldsymbol{q}(0))-1)
- \sum^{4}_{j=1} \int^{t_1}_{0} dt \left( \langle F_j(\boldsymbol{q}(t)),\tilde{q}_j(t)\rangle  + c.c.
 \right). \\ {\nonumber}
\end{eqnarray}
where $c.c.$ means complex conjugate and  $\langle \cdot,\cdot \rangle$ stands for the scalar product
\begin{equation}
\langle {a}_1,{a}_2 \rangle \equiv \int {\hat a_1}(z) {\hat a_2}^{+}(z) {\rm d} z.
\end{equation}
When ${\boldsymbol{q}}(t)$ satisfies the constraints (direct problem
plus normalization at $t=0$), all terms but the first one on the r.h.s.
of equation~(\ref{Lagrangian}) are zero and, by consequence, $L = E$
and $\delta L =\delta E $. At this stage, the adjoint
variables and the quantity $s_0$ are left unspecified. Formally the variation $\delta L$ reads as
\begin{eqnarray}
\label{equatdelta1}
\delta L=\sum^{3}_{j=1} C_j \left(  \int   q^{+}_j(z,t_1) \delta q_j(z,t_1) {\rm d} z
        -s_0 \int   q^{+}_j(z,0) \delta q_j(z,0) {\rm d} z\right)     \\ {\nonumber}
        - \sum^{4}_{j=1}\int^{t_1}_{0} dt[ \langle \delta F_j(\boldsymbol{q}(t)), \tilde{q}_j(t)\rangle
        + \langle F_j(\boldsymbol{q}(t)), \delta \tilde{q}_j(t) \rangle] + c.c..
\end{eqnarray}
 The expression $\langle F_j(\boldsymbol{q}(t)), \delta \tilde{q}_j(t) \rangle$
in equation (\ref{equatdelta1}) is zero if the governing equations
$F_j(\boldsymbol{q})=0$ are satisfied during the time interval $[0,t_1]$. The main idea then
amounts  to rewriting  quantity $\langle \delta F_j(\boldsymbol{q}(t)), \tilde{q}_j(t)\rangle$
in terms of $\delta q_k(t)$. This  is done  by  integrating by parts
in space or time. After some tedious algebra, the following identity
\begin{eqnarray}
\label{equatdelta3}
\sum^{4}_{j=1} \int^{t_1}_{0} dt \langle \delta F_j(\boldsymbol{q}(t)), \tilde{q}_j(t) \rangle=
\sum^{4}_{j=1} [\int^{t_1}_{0} dt \langle  {\tilde F}_j(\tilde{q}(t),q), \delta {q}(t) \rangle] +
\\ {\nonumber}
+ \frac{1}{Pr}\sum^{2}_{j=1}[ \int   \tilde{q}^{+}_j(z,t_1) \delta q_j(z,t_1) {\rm d} z
-   \int   \tilde{q}^{+}_j(z,0) \delta q_j(z,0) {\rm d} z]\\\nonumber
+  \int   \tilde{q}^{+}_3(z,t_1) \delta q_3(z,t_1) {\rm d} z
-   \int   \tilde{q}^{+}_3(z,0) \delta q_3(z,0) {\rm d} z
+ B(\delta\boldsymbol{q},\tilde{\boldsymbol{q}})
\end{eqnarray}
can be established, where  ${\tilde F}_j$ is an expression containing spatial or time derivatives
of $\tilde{\boldsymbol{q}}$. Note that the second, third and fourth r.h.s
terms originate from integrations by parts of time derivatives in equations
(3.8)--(3.10) and terms $B(\delta\boldsymbol{q},\tilde{\boldsymbol{q}})$ are
generated from the boundary terms resulting from integrations by parts of
spatial derivatives. These latter  terms hence  involve only quantities $\delta\boldsymbol{q}$ and
$\tilde{\boldsymbol{q}}$ evaluated at the boundaries $z=0$ and $z=1$.

\noindent At this stage,  the freedom of the Lagrangian multipliers can be used  to impose some added constraints on the adjoints fields  $\tilde{\boldsymbol{q}}$, namely: 
(i) equations ${\tilde F}_j(\tilde{\boldsymbol{q}}(t))=0, j=1 \ldots 4$, which are  similar to $F_j(\boldsymbol{q}(t))$ for  $ {\boldsymbol{q}}$ and  define 
the evolution equations (3.15)--(3.17);
and (ii) boundary conditions $B(\delta\boldsymbol{q},\tilde{\boldsymbol{q}})=0$, which  are the counterpart of boundary 
conditions (3.10)--(3.12) on ${\boldsymbol{q}}$
and define   the boundary conditions (3.18)--(3.19) for $\tilde{\boldsymbol{q}}$. This new system can  be simulated as the direct problem. It is easily seen
 that the adjoint system (3.15)--(3.19)  must be integrated backwards in time. When
it is satisfied, the variation $\delta L$ reads  
\begin{eqnarray}
\label{equatdelta4}
\delta L&=&\sum^{2}_{j=1} \left(\int (C_j q^{+}_j(z,t_1)
 - \frac{1}{Pr}\tilde{q}^{+}_j(z,t_1))\delta q_j(z,t_1) {\rm d} z\right.\\\nonumber
&& - \left.\int ( s_0  C_j q^{+}_j(z,0)-\frac{1}{Pr}\tilde{q}^{+}_j(z,0))
\delta q_j(z,0) {\rm d} z\right) \\\nonumber
&&+\int (C_3 q^{+}_3(z,t_1)
 - \tilde{q}^{+}_3(z,t_1))\delta q_3(z,t_1) {\rm d} z\\\nonumber
&& - \int ( s_0  C_3 q^{+}_3(z,0)-\tilde{q}^{+}_3(z,0))
\delta q_j(z,0) {\rm d} z+ c.c.\nonumber
\end{eqnarray}
\noindent Two relations can be still imposed, a first one  at time   $t=t_1$  which  relates $\tilde{q}_j(z,t_1)$ and ${q}_j(z,t_1) $  and a second one at  time $t=0$ which  relates $\tilde{q}_j(z,0)$ and ${q}_j(z,0) $.    These two constraints are satisfied so that  $\delta L=0$ and are  defined precisely below according to the norm and Prandtl number.
The  condition  $\delta L=0$  means that an optimal perturbation
is attained.  However this process should be self-consistent :  one   uses the  iteration procedure (3.20).
When the iterative process has converged, an initial optimal
perturbation for time $t_1$ is found.

\noindent A)  {  \bf Finite Prandtl and  zero initial temperature perturbations }

 \noindent If one  considers only  initial  perturbations in velocity  field so that variation of temperature field   
  $\delta q_3(z,0)$ is zero,  it is consistent to use the  norm $E_V$, 
i.e.,  $C_1=C_2=1$  and $C_3=0$. Equation (\ref{equatdelta4}) then naturally  leads to the relation    
\begin{equation}
\label{Endzeitrelation1}
\tilde{q}_j(z,t_1) = Pr  q_j(z,t_1),\hskip3mm j=1,2~;~~~~~~\tilde{q}_3(z,t_1) =   0
\end{equation}
at time   $t=t_1$ and the relation 
\begin{equation}
\label{Iterationsanfang1}
q_j(z,0)=\frac{1}{Pr\, s_0}\tilde{q}_j(z,0),\hskip3mm j=1,2~;~~~~~~~~ q_3(z,0)=0
\end{equation}
at time $t=0$, where $s_0$ is chosen such that the normalization condition $E(\boldsymbol{q}(0))=1$
is satisfied.

 \noindent B) {\bf { Finite Prandtl and  zero initial velocity perturbations } }
 
 \noindent When considering only  initial  perturbations in  temperature  field so that $\delta q_1(z,0)=\delta q_2(z,0)=0 $, it is consistent 
to use the  norm $E_T$, i.e.,  $C_1=C_2=0$  and $C_3=1$. Equation (\ref{equatdelta4}) then  leads to the relation     
\begin{equation}
\label{Endzeitrelation2}
\tilde{q}_j(z,t_1)=0\hskip3mm j=1,2 ~;~~~\tilde{q}_3(z,t_1) =  q_3(z,t_1)
\end{equation}
at time   $t=t_1$ and
\begin{equation}
\label{Iterationsanfang2}
q_j(z,0)=0,\hskip3mm j=1,2~;~~~~ q_3(z,0)=\frac{1}{s_0}\tilde{q}_3(z,0) 
\end{equation}
at time $t=0$
so that the normalization is satisfied.

 \noindent C) {\bf  {Infinite Prandtl number }}

 \noindent  For the infinite Prandtl number, the norm  $E_T
$  is chosen since the velocity is slaved to the temperature field in that instance,  hence $C_1=C_2=0$ and $C_3=1$. The equations are then identical  to the previous case except that only the equation for temperature appears, i.e.,

\begin{equation}
\label{EndzeitrelationInfinitePrandtl}
\tilde{q}_3(z,t_1) =  q_3(z,t_1), 
\end{equation}
and at time $t=0$
\begin{equation}
\label{IterationsanfangInfinitePrandtl}
 q_3(z,0) =\frac{1}{s_0}\tilde{q}_3(z,0)   \hskip3mm  
\end{equation}
 \noindent so that the normalization is satisfied.


\section{Numerical Method.}
\label{sec:annexB}

\noindent  For the numerical solution, the direct and adjoint equations are 
reformulated as fourth-order problem in a streamfunction-like approach. 
The incompressibility
constraint is thereby satisfied automatically, and the pressure and horizontal
velocity are eliminated
from the equations. For finite Prandtl number, the direct equations take the
form 
\begin{eqnarray}
\frac{1}{Pr} \frac{ \partial}{\partial t}  {\hat \eta}  
& = & \left[\frac{ \partial^2}{\partial z^2} - k^2  \right] {\hat \eta} - Ra k^2
 {\hat \theta},\\ 
{\hat \eta} &=& \left[\frac{ \partial^2}{\partial z^2} - k^2  \right] {\hat w}, \\  
 \frac{ \partial}{\partial t}  {\hat \theta}  
 + {\hat w}\frac{ \partial {\theta_0 } }{\partial z}   
& = & \left[\frac{ \partial^2}{\partial z^2} - k^2  \right] {\hat \theta}. 
\end{eqnarray}
The boundary conditions are 
\begin{equation}
{\hat w}= \frac{\partial {\hat w}}{\partial z} = 
\frac{\partial {\hat \theta}}{\partial z}=0  \hskip3mm\mbox{at $z=0$}
\end{equation}
and
\begin{equation}
{\hat w}=0, ~~ {\hat \eta} +k^2 Ma {\hat \theta}=0, ~~
\frac{\partial {\hat \theta}}{\partial z}+ Bi {\hat \theta} =0
\hskip3mm\mbox{at $z=1$}.
\end{equation}

 \noindent The adjoint fields satisfy the system
\begin{eqnarray}
\frac{1}{Pr} \frac{ \partial}{\partial \tau}  {\tilde \eta}  
& = & \left[\frac{ \partial^2}{\partial z^2} - k^2  \right] {\tilde \eta} - 
 k^2 {\tilde \theta}\frac{ \partial {\theta_0 } }{\partial z},\\
{\tilde \eta} &=& \left[\frac{ \partial^2}{\partial z^2} - k^2  \right] {\tilde w},\\
 \frac{ \partial}{\partial \tau}  {\hat \theta}  
& = & \left[\frac{ \partial^2}{\partial z^2} - k^2  \right] {\hat \theta} 
+Ra {\tilde w}
\end{eqnarray}
with the boundary conditions
\begin{equation}
{\tilde w}= \frac{\partial {\tilde w}}{\partial z} = 
\frac{\partial {\tilde \theta}}{\partial z}=0  \hskip3mm\mbox{at $z=0$}
\end{equation}
and
\begin{equation}
{\tilde w}= {\tilde \eta} =0, ~~
\frac{\partial {\tilde \theta}}{\partial z}+ Bi {\tilde \theta} +
Ma \frac{\partial {\tilde w}}{\partial z} =0
\hskip3mm\mbox{at $z=1$}.
\end{equation}
 \noindent  These equations are discretized in time with a backward Euler method
for the diffusive terms. The product term with
the basic temperature profile is treated with the explicit Euler method.
For the direct problem the solution at the new time level $n+1$ is
obtained by solving the following equations in sequence:
\begin{eqnarray}
\left[\frac{ \partial^2}{\partial z^2} - k^2  -\frac{1}{\Delta t}\right] 
{\hat \theta}^{n+1} 
& = & -\frac{{\hat \theta}^{n} }{\Delta t} +
 {\hat w}^n\frac{ \partial {\theta_0^n} }{\partial z},   
\\ 
\left[\frac{ \partial^2}{\partial z^2} - k^2  -\frac{1}{Pr \Delta t}\right] 
{\hat \eta}^{n+1} 
& = & -\frac{{\hat \eta}^{n} }{Pr \Delta t} +
Ra k^2  {\hat \theta}^{n+1}, \label{eta-eq-numerics}\\ 
\left[\frac{ \partial^2}{\partial z^2} - k^2  \right] {\hat w}^{n+1}
& = & {\hat \eta}^{n+1}. 
\end{eqnarray}
The boundary equations for ${\hat \eta}$ are given in terms of ${\hat w}$.
To satisfy them, the solution of the second and third equation
is represented by the linear combination
\begin{eqnarray}
{\hat \eta}^{n+1} & = & {\hat \eta}_P + \lambda {\hat \eta}_1 + 
\mu {\hat \eta}_2,\\
{\hat w}^{n+1} & = & {\hat w}_P +\lambda {\hat w}_1 + \mu {\hat w}_2,
\end{eqnarray}
where the solution with subscript $P$ is a particular solution of the 
${\hat \eta}$-equation (\ref{eta-eq-numerics}) 
with ${\hat \eta}_P=0$ on the boundaries $z=0$ and $z=1$.
The  functions with subscripts $1$ and $2$ 
satisfy the homogeneous ${\hat \eta}$-equation with zero right hand side
and two linearly independent boundary conditions, which we choose as
\begin{eqnarray}
&&{\hat \eta}_1 (z=1) ={\hat \eta}_1 (z=0) =1,\\
&&{\hat \eta}_2 (z=1) =- {\hat \eta}_2 (z=0) =1.
\end{eqnarray}
The boundary conditions $\partial {\hat w}/\partial z=0$ at $z=0$ and
${\hat \eta} + k^2 Ma {\hat \theta}=0$ at $z=1$ determine the coefficients
$\lambda$ and $\mu$ in the linear combination. We note that the functions ${\hat w}_P$,
${\hat w}_1$ and ${\hat w}_2$ satisfy zero boundary conditions at $z=0$  and
$z=1$. 

 \noindent  The adjoint solution at the new time level $n+1$ is likewise found by 
solving the equations 
\begin{eqnarray}
\left[\frac{ \partial^2}{\partial z^2} - k^2  -\frac{1}{Pr \Delta \tau}\right] 
{\tilde \eta}^{n+1} 
& = & -\frac{{\tilde \eta}^{n} }{Pr \Delta \tau} - 
k^2   {\tilde \theta}^n\frac{ \partial {\theta_0^n} }{\partial z}, \\ 
\left[\frac{ \partial^2}{\partial z^2} - k^2  \right] {\tilde w}^{n+1}
& = & {\tilde \eta}^{n+1},\\
\left[\frac{ \partial^2}{\partial z^2} - k^2  -\frac{1}{\Delta \tau}\right] 
{\tilde \theta}^{n+1} 
& = & -\frac{{\tilde \theta}^{n} }{\Delta \tau} -Ra {\tilde w}^{n+1}
\end{eqnarray}
in the given sequence.
The solution for ${\tilde w}^{n+1}$ and ${\tilde \eta}^{n+1}$ must again be represented
as a linear combination with auxiliary functions satisfying the homogeneous
$\tilde \eta$-equation in order to satisfy the boundary conditions.

 \noindent  Discretization in space is realized with an expansion in Chebyshev 
 polynomials (see \cite{Canuto:1988}).
Product terms with the perturbation and the basic state are calculated 
in physical space by a fast cosine transform. 
The  Helmholtz equations for the variables 
$\hat \eta$, $\hat w$, $\hat \theta$ and the adjoint variables
$\tilde \eta$, $\tilde w$, $\tilde \theta$ reduce to  essentially 
tridiagonal linear systems. 
The boundary conditions 
are treated with the tau method, which produces two filled rows in the matrix
representation. The limit of infinite Prandtl number requires
no changes in the solution procedure.

\noindent The basic temperature profile is computed  with the same numerical
method as the perturbations, i.e. using the  backward-Euler method 
and a Chebyshev polynomial expansion with the same time step  and number of
polynomials.  The entire field $\theta_0(z,t)$ is 
stored in an  array in order to speed up the  backward integration  of the adjoint equations.

\noindent  The code was tested for infinite Prandtl number with a stationary basic 
temperature profile. It was verified that
exponential growth of the optimal perturbations appeared at the proper threshold
values of $Ma\approx 79.6$ for pure Marangoni convection with $Bi=0$ 
\cite[]{Pearson:1958}
 and
for $Ra\approx 1100$ for pure Rayleigh convection with fixed temperature on the
free surface \cite[]{Chandrasekhar:1961}
 For this verification, the boundary 
condition at the bottom was changed to constant temperature.


\section{Analysis of Stability for the Marangoni case }
\label{AnnexC}

\subsection{ The approach  for the Marangoni case~($Ra=0$).}
\label{approach}

\noindent  This section presents an  approach  valid for infinite Prandtl number,  
which evaluates  the evolution in terms of orders of magnitude.  It is based on two hypotheses which make the analysis tractable. 
First the initial perturbation of wavenumber $k$ in the $x$ direction   is   only a temperature perturbation i.e.
$ {\hat u}(z,t=0)={\hat w}(z,t=0)=0$ and the temperature perturbation ${\hat \theta}(z,t=0)$  
is  uniform along  the $z$-direction.    
Second, the flow is supposed unstable  i.e. convection sets in,  if  there exists   a  time and  a region in the  flow in which   the advection term in  
equation (3.10)  becomes greater  or equal to the two diffusion terms  which tend to damp  the initial perturbation.

\noindent Practically, one first   determines  the orders of
magnitude  for   temperature  perturbations
 when the system  is
in a stable regime or near the critical curve, i.e., when the term corresponding to
advected heat transfer  ${\hat w}\frac{ \partial {\theta_0 }
}{\partial z} $     in equation (3.10) can be neglected,
according to the second hypothesis.
Thereafter one computes the order of magnitude   of the term $k^2
{\hat \theta }$
i.e., the diffusion  in  the $x$-direction,  
and  of  the term   $\frac{ \partial^2  {\hat \theta }}{\partial z^2}$, i.e.,
the diffusion in  the $z$-direction, corresponding to the stable
regime. 
This is done in subsection \ref{temperatureScaling}. 
On the other hand, the order of magnitude  for velocity  ${\hat w}$
is found in  subsection  \ref{velocityBM}, as well as the
corresponding advection term,  ${\hat w}\frac{ \partial {\theta_0
} }{\partial z}  $. 

\noindent Since the velocity is computed using the
temperature perturbation field estimated for the stable
configuration, this approach is consistent only if the advection
term remains much smaller than one of the diffusion terms. The  sets  of
parameters $Ma,Bi,Pr$ that give consistent results for each time
and mode are considered in the "stable" domain. Otherwise, if
there exists a time and a mode of
wavenumber  $k$
such that   the advection term  is larger  in order of magnitude
than the two diffusion terms, the corresponding set of parameters
is associated with a situation where convection sets in
(subsection~\ref{ONSETSCALINGMarangoni}). 
The scaling laws for critical parameters are then derived by solving
the resulting set of inequalities (subsection~\ref{SOLVEINEQMarangoni}).

\subsection{ Scaling Analysis for the Temperature Perturbation Field}
\label{temperatureScaling}

\noindent  As mentioned in the previous paragraph, we  determine   the orders of magnitude  for   temperature  perturbations by  neglecting
the advected heat transfer  ${\hat w}\frac{ \partial {\theta_0 } }{\partial z}  $     in equation (3.10).     One thus obtains  
 \begin{equation}
 \label{Direct4NEW}
 \frac{ \partial {\hat \theta} }{\partial t}     - \left[\frac{ \partial^2}{\partial z^2} - k^2  \right]{ {\hat \theta} }=0
 \end{equation} 
The temperature perturbation field also  satisfies the boundary conditions 
\begin{eqnarray}
\label{pertTemperatureScaling1}
\frac{\partial {\hat \theta}}{\partial z} +Bi\,  {\hat \theta} =0,    \hskip3mm\mbox{at $z=1$},\\
\label{pertTemperatureScaling2}
\frac{\partial {\hat \theta}}{\partial z}=0  \hskip3mm\mbox{at $z=0$}.
\end{eqnarray}

\noindent   The cooling due to evaporation imposes  that a thermal boundary layer for the temperature perturbation ${\hat \theta}(z,t)$  
 develops.  One can easily check that  the solution 
 \begin{equation}
 \label{solutiontemepraturepert}
  {\hat \theta}(z,t)=a_0(1+{\theta}_{0}(z,t))\exp(-k^2t)
 \end{equation} 
 satisfies the above system  and the condition of uniformity at $t=0$. Note that  $a_0$ is simply the initial  amplitude of the temperature perturbation which is taken to be equal to one in the sequel.  
From the above equation, it is readily seen that   the   thickness  of  the thermal boundary layer    for the   perturbation field  ${\hat \theta}(t)$ is  equal to 
$\delta_{0}(t)$ and that the scale of   variation in the $z$-direction of  the perturbed  field ${\hat \theta}(z,t)$ denoted by   $ \Delta {\hat \theta} $  satisfies
 \begin{equation}
\label{A1}
\Delta {\hat \theta} (t)  \sim  \Delta {\theta}_{0}(t) \exp(-k^2t)
\end{equation}
The scale $  {\hat \theta}_s $ of     the perturbed  field ${\hat \theta}(z=1,t)$  on the surface    satisfies according to equation~(\ref{pertTemperatureScaling1})
 \begin{equation}
\label{A6}
{\hat \theta}_s \sim  \frac{1 }{Bi }   \frac{\Delta {\hat \theta} (t)  }{\delta_{0}(t) }
\end{equation}

\noindent Using  the  results of Annex~A, it is straightforward to find the following estimates:

\noindent For  $Bi \ll  1$ :
\begin{equation}
\label{A2}
\Delta {\hat \theta} (t)  \sim    Bi\sqrt{t} \exp(-k^2t) ,~~~~ {\hat \theta}_s  \sim  \exp(-k^2t), ~~   {\hat \theta}   \sim  \exp(-k^2t)~~~~\hbox{for }~~0 \lesssim \sqrt{t} \lesssim  1,
\end{equation}
\begin{equation}
\label{A3}
~~\Delta {\hat \theta} (t)  \sim    Bi  \exp(-k^2t) ,~~~~ {\hat \theta}_s  \sim  \exp(-k^2t) , ~~   {\hat \theta}   \sim  \exp(-k^2t)      ~~~~\hbox{for }~~  1 \lesssim  t \lesssim  Bi^{-1}.
\end{equation}

\noindent For   $  1  \ll Bi $ : 
 \begin{equation}
\label{A4}
\Delta {\hat \theta} (t)  \sim    Bi\sqrt{t} \exp(-k^2t) ,~~~ {\hat \theta}_s  \sim  \exp(-k^2t),   ~~   {\hat \theta}   \sim  \exp(-k^2t)~~~\hbox{for }~~0 \lesssim {t} \lesssim   Bi^{-2},
\end{equation}
\begin{equation}
\label{A5}
\Delta {\hat \theta} (t)  \sim   \exp(-k^2t) ,   ~~   {\hat \theta}_s   \sim  \frac{ \exp(-k^2t) }{Bi \sqrt{t}}, ~~~~ {\hat \theta}   \sim  \exp(-k^2t),~\hbox{for }~~ Bi^{-2}  \lesssim  {t}  \lesssim  1.
\end{equation}


\subsection{ Scalings  for the  Velocity  Field in the B\'enard-Marangoni problem ($Ra=0$).}
\label{velocityBM}

\noindent The equation of the  vorticity field can be easily deduced from equations (3.8) and  (3.9).   Denoting by 
 ${\hat \omega}$ the $y$-component of vorticity, we obtain the diffusion equation
\begin{equation}
\label{Vorticity}
        \frac{1}{Pr} \partial_t {\hat \omega} =  \partial^2_z {\hat \omega} -   k^2 {\hat \omega}
\end{equation}
\noindent   For infinite Prandtl number,  this equations simplifies 
\begin{equation}
\label{Vorticity11}
   \partial^2_z {\hat \omega} -   k^2 {\hat \omega}=0
\end{equation}
\noindent The vorticity   is slaved to the temperature evolution  {\it via} the boundary condition at the free surface given by equation (3.11) :
\begin{equation}
\label{VorticityBCtop}
        {\hat \omega} +   ik Ma {\hat \theta}=    0~~~  \hbox{at}~~~z=1.
\end{equation}

\noindent   Equation (\ref{Vorticity11})    plus the forcing (\ref{VorticityBCtop})  defines an hydrodynamic  boundary layer $ \delta_H $. It is easily seen that the proper scaling reads  
\begin{equation}
\label{HydroEvolution}
 \delta_H \sim \min(  1/k , 1)
\end{equation}
\noindent The hydrodynamic  boundary layer either reaches the bottom,  i.e., $\delta_H \sim 1$, or 
the  diffusion term along $x$  becomes of the same order of  the diffusion term  along $z$ and  $ \delta_H   \sim 1/k$.

\noindent From  equation (3.11),  a relation between ${\hat \theta}_{s}$ and  the order of magnitude of   velocity
$  {\hat u} $   can be found   :

\begin{equation}
\label{HydroEvolution1}
 {\hat u}  \sim k  \delta_H ~ Ma ~{\hat \theta}_{s}(t).
\end{equation}
  \noindent The order of magnitude of  the  vertical component ${\hat w}$ of  velocity  is provided   { \it via } mass conservation

 \begin{equation}
\label{HydroEvol1}
{\hat w} \sim    k \delta_{H}   {\hat u}  \equiv  ( k  \delta_H)^2 ~ Ma ~{\hat \theta}_{s}(t).
\end{equation}


\subsection{Condition for the onset of convection for the Marangoni flow ($Ra=0$) }
\label{ONSETSCALINGMarangoni}

\noindent  To describe  the time  evolution of  perturbations,  one must    distinguish  two  regions along the  $z$ direction i.e.  inside and outside the thermal boundary layer.
Outside the layer ($ \delta_{0}  \lesssim  1-z   \lesssim 1$),  the advection term in equation (3.10)  is zero since the basic  temperature field $  \theta_{0}(z,t)$ vanishes : hence diffusion    dominates and perturbations are always damped. Instability thus only arises inside the thermal layer  ($ 0  \lesssim   1- z \lesssim   \delta_{0} $).

\noindent To determine the onset of instability, one  first  compares   the order of magnitude  of  the advection   and  the diffusion along   the $x$-direction  in the thermal layer  
\begin{equation}
\label{OnsetCriterionX}
\frac{ {\hat w}_{th}  \Delta {\theta_0} }{\delta_{0} }  \; , \; k^2 {\hat \theta} \; 
\end{equation}
 \noindent  where  $  {\hat w}_{th}$ denotes    the order of magnitude   of the vertical velocity  in the thermal boundary layer,  and second  
 the order of magnitude  of  the advection term  and of  the diffusion term in  the $z$-direction  
\begin{equation}
\label{OnsetCriterionZ}
\frac{ {\hat w}_{th}\Delta {\theta_0} }{\delta_{0} }  \; , \;  \frac{   \Delta {\hat \theta}} { \delta_{0}^2  }  \; 
\end{equation}
 The  existence of a  convection onset   thus implies that there exists a time and
  a mode of wavenumber  $k$ for which  the two conditions 
   \begin{equation}
\label{OnsetCriterion2a}
 {\hat w}_{th} ( \frac{ \Delta {\theta_0} }{\delta_{0} })      \gtrsim     k^2 {\hat \theta}~~~     \hbox{and}  
  ~~~{\hat w}_{th} ( \frac{ \Delta {\theta_0} }{\delta_{0} })   \gtrsim       \frac{   \Delta {\hat \theta}} { \delta_{0}^2  }    
\end{equation}
 hold.  We need  quantity $  {\hat w}_{th}$  since it explicitely appears in the above inequalities.
Two possibilities  should be considered  at each time : $\delta_0(t)  \lesssim \delta_H  $     or    $ \delta_H  \lesssim  \delta_0(t)$. In the first case,  the thermal layer is included in the 
hydrodynamic  layer and one may use the scaling $  {\hat w}_{th} \sim   \frac{\delta_0}{\delta_H}  {\hat w} $. In  the second case   $  {\hat w}_{th} \sim   {\hat w} $.  
Using equation~(\ref{HydroEvol1}), this implies that  the scaling  $  {\hat w}_{th}$    is such that
 \begin{eqnarray}
\label{HydroReg1}
 {\hat w}_{th}   & \sim &  \min\left(1, \frac{\delta_0}{\delta_H}\right) {\hat w} =   \min\left(1, \frac{\delta_0}{\delta_H}\right) ( k  \delta_H)^2 ~ Ma ~{\hat \theta}_{s}.
\end{eqnarray}

\noindent  It  is  now possible to rewrite inequalities~(\ref{OnsetCriterion2a})  as 
 \begin{equation}
\label{OnsetCriterion2aB}
\min\left(1, \frac{\delta_0}{\delta_H}\right) \delta_H^2 ~ Ma ~{\hat \theta}_{s} ( \frac{ \Delta {\theta_0} }{\delta_{0} })      \gtrsim     {\hat \theta}~~~     \hbox{and}  
  ~~~\min\left(1, \frac{\delta_0}{\delta_H}\right) ( k  \delta_H)^2 ~ Ma ~{\hat \theta}_{s} \Delta {\theta_0}  \gtrsim       \frac{   \Delta {\hat \theta}} { \delta_{0}  }    
\end{equation}

\subsection{Derivation of scaling laws}
\label{SOLVEINEQMarangoni}

\noindent One must now   introduce the  various expressions previously obtained  for   $\Delta {\theta_0} $,  $ \delta_{0}$,  $  {\hat \theta}$,  $ \Delta {\hat \theta}$, $  {\hat \theta}_s$,  $\delta_H$    
 inside  instability conditions~\ref{OnsetCriterion2aB}.
The  expressions for  $ \delta_{0}$ and  $\Delta {\theta_0} $ are  obtained  in section~(\ref{sec:BasicState}), 
the  expressions for  ${\hat \theta}$,  $ \Delta {\hat \theta}$ and  $  {\hat \theta}_s$ in section~(\ref{temperatureScaling}), the  expression for  $ \delta_{H}$   in section~(\ref{velocityBM}).
 To ease the discussion,     three separate  cases are studied :  
 
 \noindent  A) $Bi \ll  1$ ,~~~ B)  $1 \ll  Bi$ ~and~ $ t   \lesssim   Bi^{-2}$,~~~ C)   
 $1 \ll Bi$ ~and  $Bi^{-2} \lesssim t$.


\subsubsection*{ {\bf   A) Case} $Bi \ll  1$ }

\noindent First let us recall   from  Annex~A,  that   the following relations hold~:
\begin{equation}
\label{BM1}
\delta_{0}(t)  \sim \min(\sqrt t,1)   ~~~ \hbox{and} ~~~   \frac{\Delta {\theta_0}(t) }{\delta_{0}(t)} \sim  Bi.
\end{equation}

\noindent  From  section~\ref{temperatureScaling}, one easily verifies that the temperature perturbation field is such that  
\begin{equation}
{\Delta {\hat \theta}}  \sim Bi {\delta_0} {\hat \theta}_{s}~~~ \hbox{and} ~~~{\hat \theta} \sim  {\hat \theta}_{s} 
 \label{BM2} 
\end{equation}

  \noindent Using  equations~(\ref{BM1}) and  (\ref{BM2}), condition   (\ref{OnsetCriterion2aB}) can be transformed  into 
\begin{equation}
\label{BM3}
~\min\left(1, \frac{\delta_0 }{\delta_H}\right) ~Ma ~ Bi   ~    \delta_H^2 ~    \gtrsim 1 ~~~ 
\hbox{and}~~~\min\left(1, \frac{\delta_0 }{\delta_H}\right)~ \delta_0   ~ Ma ~ k^2 ~  \delta_H^2  ~    \gtrsim 1 
\end{equation}

\noindent Note that only the period $t \lesssim  1/Bi$ should be considered  here since,  for $ 1/Bi \lesssim  t$,  the basic state has relaxed to a uniform  temperature.

\noindent To ease the discussion,    two separate cases must be considered 
for the wavenumber $k$, namely     $ k  \lesssim 1 $ and $ 1  \lesssim k  $.

 \noindent $\bullet$ $ k  \lesssim 1 $ 

\noindent  In that instance,  $\delta_H \sim \min(  1/k , 1)=1$ (see equation  (\ref{HydroEvolution}))  leading to the equality 
 $\min(1, \frac{\delta_0}{\delta_H})  \sim \delta_{0}$. Condition   (\ref{BM3}) reads 
\begin{equation}
\label{BM8}
~ \delta_0 ~Ma ~ Bi  ~    \gtrsim 1 ~~~ \hbox{and}~~ \delta_0^2 ~  Ma ~ k^2 ~     \gtrsim 1 
\end{equation}
 The smallest  Marangoni  number i.e.  the critical Marangoni number which satisfies such inequalities,  is obtained  for  $\delta_0(t)  \sim  1$ i.e. for $ 1 \lesssim t $. Condition  (\ref{BM8})  becomes  
 \begin{equation}
\label{BM9}
 Ma ~     \gtrsim  \frac{1}{Bi}  ~~~ \hbox{and}~~~ k^2 ~    \gtrsim  \frac{1}{Ma} 
\end{equation}
From the above onditions,  one easily gets  the critical value  
  \begin{equation}
\label{BM10}
   Ma_c  \sim   1/Bi,    ~~~~ \hbox{with}~~~\sqrt{Bi}  \lesssim  k_{c} \lesssim 1  ~~~ \hbox{and}~~~ 1   \lesssim    t_{c}  \lesssim  1/Bi.    
 \end{equation}

\noindent $\bullet$ $ 1  \lesssim k  $. 

\noindent   In that instance,  $\delta_H \sim \min(  1/k , 1)=1/k$.  Since the two functions  $\min(1,k{\delta_0 }) $  and    $\min(1,k{\delta_0 })   \delta_{0}$
 are both increasing functions of   $\delta_{0}$ when $ \delta_{0}  \lesssim 1/k$, the critical Marangoni must be  obtained  when  $1/k  \lesssim \delta_{0}$ for which 
$\min (1,k{\delta_0 } ) = 1$. Conditions   (\ref{BM3})   become
 \begin{equation}
\label{BM6}
 Ma ~Bi  ~    \gtrsim k^2   ~~~ \hbox{and}~~~ Ma ~\delta_0(t)~    \gtrsim  1
 \end{equation}
  \noindent A straigthforward discussion directly leads to the conditions 
  \begin{equation}
\label{BM7}
Ma_c  \sim   1/Bi    ~~~~ \hbox{and}~~~  k_{c} \sim 1  ~~~ \hbox{and}~~~ 1   \lesssim    t_{c}   \lesssim  1/Bi.   
 \end{equation}
which is a limiting case of  condition (\ref{BM10}).   As a consequence, the critical conditions  in the case $Bi \ll 1$ corresponds to condition (\ref{BM10}).


 \subsubsection*{  {\bf   B) Case}  ~~$ 1 \ll Bi$ ~~and~~ $ t   \lesssim   Bi^{-2}$.  }

 \noindent From   results obtained on the basic flow,  it is easily seen that  relations (\ref{BM1}),  (\ref{BM2}) and thus  (\ref{BM3})   still hold. Moreover 
note that   the largest  value of  $\delta_0(t)$  is obtained at the largest time $t \sim  Bi^{-2}$:   $\delta_0(Bi^{-2}) \sim Bi^{-1}$. 
   One must  consider  the three cases    $ k  \lesssim 1 $, ~~ $ 1  \lesssim k   \lesssim Bi $~~ and~~  $Bi~\lesssim~k $.

 \noindent $\bullet$ $ k  \lesssim 1 $ 
 
  \noindent  In that case,  $\delta_H \sim \min(  1/k , 1)=1$ and  condition   (\ref{BM8}) is again satisfied. 
  The critical Marangoni number, is obtained  for the largest possible  value of  $\delta_0(t)$  i.e.  $\delta_0(Bi^{-2})= Bi^{-1}$. 
Condition~(\ref{BM8}) now reads  
 \begin{equation}
\label{BM15}
 Ma ~     \gtrsim 1 ~~~ \hbox{and}~~ Ma~ k^2 ~    \gtrsim  Bi^{2}
\end{equation}
 
  \noindent  A straigthforward discussion directly leads to the   critical  conditions for instability     
 \begin{equation}
\label{BM13}
Ma_c \sim   Bi^2  ~~\hbox{with}~~~   k_{c}   \sim  1~~ \hbox{and}~~~   t_{c}  \sim   Bi^{-2}
 \end{equation}

 \noindent $\bullet$ $ 1  \lesssim k   \lesssim Bi $ 
 
\noindent  In that instance,  $\delta_H \sim \min(  1/k , 1)=1/k$. Conditions   (\ref{BM3}) become
\begin{equation}
\label{BM16}
~\min(1,k{\delta_0(t)}) ~    k^{-2} ~Ma ~ Bi    ~    \gtrsim 1 ~~~ \hbox{and}~~~\min(1, k{\delta_0(t)})~ \delta_0(t)   ~ Ma~    \gtrsim 1 
\end{equation}
\noindent  The   critical Marangoni number, is obtained  for the largest possible  value of  
  $\delta_0(t)$ obtained at the largest time i.e. $t=Bi^{-2}$  for which  $\delta_0(Bi^{-2})= Bi^{-1}$. As a consequence 
  \begin{equation}
\label{BM17}
~\min (1, k/Bi) ~    k^{-2} ~Ma ~ Bi    ~    \gtrsim 1 ~~~ \hbox{and}~~~\min (1, k/Bi)~  Bi^{-1}   ~ Ma~    \gtrsim 1 
\end{equation}
For this wavenumber interval, $  \min (1, k/Bi) =k/Bi$ and equation~(\ref{BM17}) becomes 
 \begin{equation}
\label{BM18}
  k^{-1} ~Ma      ~    \gtrsim 1 ~~~ \hbox{and}~~~ k~ Bi^{-2}    ~ Ma~    \gtrsim 1 
\end{equation}

 \noindent   This leads  to the critical  conditions for instability     
 \begin{equation}
\label{BM14}
Ma_c \sim   Bi   ~~\hbox{with}~~~   k_{c}   \sim  Bi~~ \hbox{and}~~~   t_{c}  \sim   Bi^{-2} 
 \end{equation}

 \noindent $\bullet$  $  Bi  \lesssim k     $
   
 \noindent  In that case,  $\delta_H \sim \min(  1/k , 1)=1/k$ and  conditions~(\ref{BM16}) are verified. Again  
 the   critical Marangoni number, is obtained  for the largest  time i.e. $t=Bi^{-2}$ corresponding to the largest possible  value of 
 $\delta_0(t)$.  Moreover, for this wavenumber interval, $  \min (1, k{\delta_0(t)}) =1$  and equation~(\ref{BM16}) becomes 
 \begin{equation}
\label{BM19}
     k^{-2} ~Ma ~ Bi    ~    \gtrsim 1 ~~~ \hbox{and}~~~  Bi^{-1}   ~ Ma~    \gtrsim 1 
\end{equation}

  \noindent   This leads  to  the same critical  conditions~(\ref{BM14}) for instability:
  
   \begin{equation}
\label{BM14bis}
Ma_c \sim   Bi   ~~\hbox{with}~~~   k_{c}   \sim  Bi~~ \hbox{and}~~~   t_{c}  \sim   Bi^{-2}  
 \end{equation}


 \subsubsection*{  {\bf   C) Case} ~~$ 1 \ll Bi$ ~~and ~~$ Bi^{-2}  \lesssim t \lesssim1 $  }

First let us recall   from  results on the basic flow  that   the following relation holds:
\begin{equation}
\label{BM11}
\delta_{0}(t)  \sim  \sqrt t  ~~~ \hbox{and} ~~~  {\Delta {\theta_0}(t) }  \sim      {1 }
\end{equation}

\noindent  From  section~\ref{temperatureScaling}, one easily verifies that the temperature perturbation field is such that  
\begin{equation}
{\Delta {\hat \theta}}  \sim  Bi {\delta_0} {\hat \theta}_{s} ~~ \hbox{and}~{\hat \theta}  \sim  \Delta {\hat \theta}
 \label{BM12} 
\end{equation}

  \noindent  Using equations~(\ref{BM11}), (\ref{BM12}),    
conditions~
(\ref{OnsetCriterion2aB})
can be transformed  into 
\begin{equation}
\label{BM20}
~\min\left(1, \frac{\delta_0}{\delta_H}\right) ~ \delta_0^{-2}~    \delta_H^2 ~  Ma ~ Bi^{-1}       \gtrsim 1 ~~~ \hbox{and}~~~\min\left(1, \frac{\delta_0}{\delta_H}\right)~  \delta_H^2  ~ k^2  ~  Ma ~ Bi^{-1}    \gtrsim 1 
\end{equation}
 
\noindent  At this stage,    two possibilites should be considered~:   $ k  \lesssim 1 $ ~~ and ~~$  1  \lesssim k     $.


  \noindent $\bullet$ $ k  \lesssim 1 $ 
 
 \noindent  In that case,  $\delta_H \sim \min(  1/k , 1)=1$ and $\min (1, \frac{\delta_0}{\delta_H})  \sim \delta_{0}$.  Conditions~(\ref{BM20}) become 
 \begin{equation}
\label{BM21}
\delta_0^{-1}~ ~  Ma ~ Bi^{-1}       \gtrsim 1 ~~~ \hbox{and}~~~  \delta_0  ~ k^2  ~  Ma ~ Bi^{-1}    \gtrsim 1.
\end{equation}
    
   \noindent By multiplying both  conditions, one gets: 
   \begin{equation}
\label{BM27}
  Ma ~         \gtrsim~  Bi~  k^{-1}.  
 \end{equation}
   \noindent This implies that $ k_{c}   \sim  1$ and $Ma_c \sim   Bi$. Introducing the latter two equalities back  into equation (\ref{BM21})
provides $ \delta_0 =1$ i.e. $ t_{c}  \sim   1$.  Finally the critical  conditions can be written as
 \begin{equation}
\label{BM22}
Ma_c \sim   Bi   ~~\hbox{with}~~~   k_{c}   \sim  1~~ \hbox{and}~~~   t_{c}  \sim   1.
 \end{equation}

 \noindent $\bullet$ $ 1  \lesssim k  $

 \noindent  In that instance,  $\delta_H \sim \min(  1/k , 1)=1/k$.  If one introduces the new variable  $\xi \equiv k{\delta_0} $, conditions   (\ref{BM20})  read 

\begin{equation}
\label{BM23}
Ma   \gtrsim Bi F(\xi)~~~ \hbox{and}~~~Ma   \gtrsim Bi G(\xi)
\end{equation}
 \noindent  in which 
  \begin{equation}
\label{BM25}
F(\xi) \equiv \xi^2 \min(1,\xi)^{-1}  ~~~ \hbox{with}~~~G(\xi) \equiv \min(1,\xi)^{-1}.  
\end{equation}
 
 \noindent   A straightforward  analysis of these two functions shows that the critical Marangoni is reached for $\xi=1$ hence $Ma_c \sim   Bi$. 
 Moreover, since  $ 1/Bi  \lesssim\delta_0   \lesssim 1$ in this time interval,  a large bandwith of modes $k$ are equivalent 
  leading to the following critical  conditions for instability:  
\begin{equation}
\label{BM28}
Ma_c \sim   Bi  ~~  \hbox{with}~~~1 \lesssim   k_{c}   \lesssim   Bi ~~ \hbox{and}~~~      t_{c}   \sim   k_{c}^{-2} 
 \end{equation}


 \noindent   Finally, by taking the lowest Marangoni numbers   of the  conditions (\ref{BM13})-(\ref{BM14})-(\ref{BM22})-(\ref{BM28}), one 
  deduces  the true critical conditions    for $   1 \ll Bi $, namely
\begin{equation}
\label{final2}
Ma_c \sim   Bi  ~~  \hbox{with}~~~1 \lesssim   k_{c}   \lesssim   Bi ~~ \hbox{and}~~~      t_{c}   \sim   k_c^{-2}.
 \end{equation}

\newpage


\begin{figure}
\scriptsize{
\centerline{
\includegraphics[width=0.45\textwidth,clip=]{pics/macvsbi.eps}
\hfill \includegraphics[width=0.45\textwidth,clip=]{pics/alphacvsbi-mapure.eps}}
}
\centerline{
\includegraphics[width=0.45\textwidth,clip=]{pics/tcvsbi-mapure.eps}
\hfill
}
\caption{Infinite Prandtl number case and $Ra=0$. Results are shown for two thresholds $G_{thres}=1$ and $G_{thres}= 100$. 
Frozen-time and steady state results are presented for comparison (see text for details).
(a) Critical Marangoni $Ma_c(Bi)$, (b) Critical wavenumber $k_c(Bi)$, (c) Critical time $t_c(Bi)$.
}
\label{CriticalMarangoniPrINF}
\end{figure}

\begin{figure}
\scriptsize{
\centerline{
\includegraphics[width=0.45\textwidth,clip=]{pics/racvsbi.eps}
\hfill\includegraphics[width=0.45\textwidth,clip=]{pics/alphacvsbi-rapure.eps}}
}
\centerline{
\includegraphics[width=0.45\textwidth,clip=]{pics/tcvsbi-rapure.eps}
\hfill
}
\caption{Infinite Prandtl number case and $Ma=0$. Results are shown for two thresholds $G_{thres}=1$ and $G_{thres}= 100$. 
Frozen-time and steady state results are presented for comparison (see text for details). 
(a) Critical  Rayleigh $Ra_c(Bi)$, (b) Critical wavenumber $k_c(Bi)$, (c) Critical time $t_c(Bi)$.
}
\label{CriticalRayleighPrINF}
\end{figure}

\clearpage 

\bibliographystyle{jfm}


\end{document}